\pdfoutput=1
\documentclass[a4paper,fleqn,usenatbib]{mnras}

\usepackage{newtxtext,newtxmath}
\usepackage[T1]{fontenc}
\usepackage{ae,aecompl}

\usepackage{amsmath, latexsym, amssymb, graphicx, slashed, color}
\usepackage{hyperref}

\definecolor{nicered}{rgb}{.7,.1,.1}
\definecolor{nicegreen}{rgb}{.1,.5,.1}
\definecolor{darkblue}{rgb}{0,0,.5}
\hypersetup{colorlinks, citecolor=nicegreen,linkcolor=nicered, urlcolor=darkblue}

\newcommand{\beq}{\begin{equation}}
\newcommand{\eeq}{\end{equation}}

\newcommand{\be}{\begin{equation}}
\newcommand{\ee}{\end{equation}}
\newcommand{\bea}{\begin{eqnarray}}
\newcommand{\eea}{\end{eqnarray}}

\newcommand{\bw}{\begin{widetext}}
\newcommand{\ew}{\end{widetext}}

\title[The universal rotation curve of LSBs]{The universal rotation curve of low surface brightness galaxies - IV.
The interrelation between dark and luminous matter}

\author[C. Di Paolo et al.]{
Chiara Di Paolo,$^{1, 2, 3}$\thanks{chiara.dipaolo@sissa.it}
Paolo Salucci,$^{1,2, 3}$\thanks{salucci@sissa.it}
Adnan Erkurt,$^{4}$
\\
$^{1}$SISSA/ISAS, Via Bonomea 265, 34136 Trieste, Italy\\
$^{2}$INFN Sez.\ Trieste, Via A. Valerio 2, 34127 Trieste, Italy\\
$^{3}$IFPU, Via Beirut 2-4, 34151 Trieste, Italy\\
$^{4}$Istanbul University, Beyazit, 34452 Fatih/Istanbul, Turkey
}

\begin{document}
\label{firstpage}
\pagerange{\pageref{firstpage}--\pageref{lastpage}}

\maketitle

\begin{abstract}
\noindent
We investigate the properties of the baryonic and the dark matter components in low surface brightness (LSB) disc galaxies, with central surface 
brightness in the B band $\mu_0 \geq 23 \, mag \, arcsec^{-2}$. 
The sample is composed by 72 objects, whose rotation curves show an orderly trend reflecting the idea of a universal rotation curve (URC) 
similar to that found in the local high surface brightness (HSB) spirals in previous works. 
This curve relies on the mass modelling of the coadded rotation curves, 
involving the contribution from an exponential stellar disc and a Burkert cored dark matter halo. We find that the dark matter is dominant especially within the smallest and less luminous LSB galaxies. 
Dark matter halos have a central surface density $\Sigma _0 \sim 100 \, M_{\odot} pc^{-2}$, similar to galaxies of different Hubble types and luminosities.
We find various scaling relations among the LSBs structural properties which turn out to be similar but not identical to what has been 
found in HSB spirals. In addition, the investigation of these objects calls for the introduction of a new luminous parameter, the stellar compactness $C_*$ (analogously to a recent work by Karukes \& Salucci), alongside with the optical radius and the optical velocity in order to reproduce the URC. Furthermore, a mysterious entanglement between the properties of the luminous and the dark matter emerges.
 
\leavevmode
\null

\end{abstract}

\makeatletter
\@thanks
 \gdef\@thanks{}
 \SFB@keywordstrue
\makeatother

\section{Introduction}\label{introduction}

\noindent
Dark matter (DM) is a main actor in cosmology. It is believed to constitute the great majority of the mass and to rule the processes of structure formation in the Universe. \footnote{In this paper we adopt the scenario of DM in Newtonian gravity, leaving to other works the investigation in different frameworks.} The so-called $\Lambda$CDM ($\Lambda$ Cold Dark Matter) scenario, in which one assumes a WIMP (Weakly Interactive Massive Particle) that decouples from the primordial plasma when non-relativistic, successfully reproduces the structure of the cosmos on large scales ~\citep{Kolb_Turner_1990}.
However, some challenges to this scenario emerge at small galactic scales, such as the 'missing satellite problem' (e.g ~\citealp{Klypin_1999, Moore_1999,  Zavala_2009,Papastergis_2011, Klypin_2015}) and the 'too big to fail problem' (e.g. ~\citealp{Ferrero_2012, Boylan_2012, Garrison_Kimmel_2014, Papastergis_2015}).
Moreover, the galactic inner DM density profiles generally appear to be cored, rather than cuspy as predicted in the $\Lambda$CDM scenario (e.g. \citealp{Salucci_2001, deBlok_2002, Gentile_2004, Gentile_2005,Simon_2005, Del_Popolo_2009, Oh_2011,Weinberg_2013}), 
in spirals of any luminosity (see \citealp{Salucci_2019}). In ellipticals and dSphs the question is still uncertain ( \citealp{Salucci_2019}). 
\begin{figure*}
\begin{center} 
\includegraphics[width=1.05\textwidth,angle=0,clip=true]{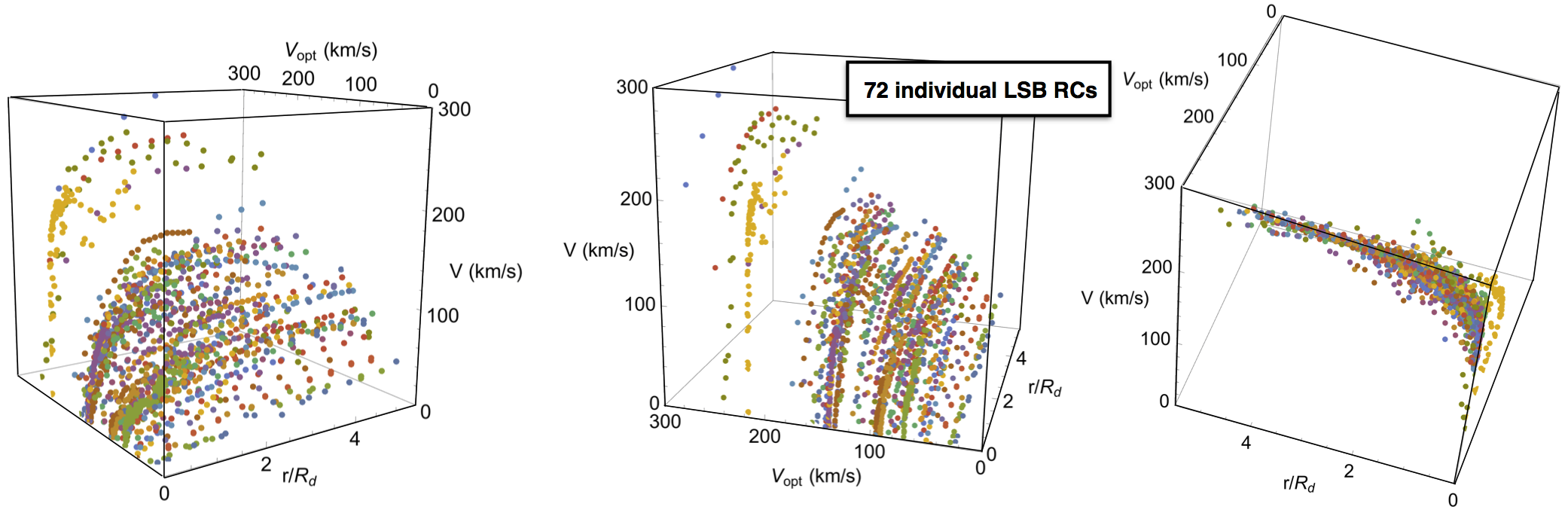} 
\caption{LSBs rotation curves (each one in different color) ordered according to increasing optical velocities $V_{opt}$. Note that the radial coordinate is normalised with respect to the disc scale length $R_d$. A {\it universal trend} is recognizable analogous to that emerged in  normal spirals (see Fig. \ref{RC_Salucci} in Appendix \ref{RC_Normal_Spirals}).}  
\label{RCs_3D}
\end{center}
\end{figure*}

These issues suggest to study different scenarios from  the 'simple' $\Lambda$CDM, such as warm DM (e.g. ~\citealp{de_Vega_2013, Lovell_2014}), self-interacting DM (e.g ~\citealp{Vogelsberger_2014, Elbert_2015}), or to introduce the effect of the 
baryonic matter feedbacks on the DM distribution (e.g. ~\citealp{Navarrot_1996, Read_2005, Mashchenko_2006, Pontzen_2014, Di_Cintio_2014}). 
\\
\\
One important way to investigate the properties of DM in galaxies is to study rotational supported systems, such as spiral galaxies, since they have a rather simple kinematics. 
The stars are mainly distributed in an exponential thin disc with scale length $R_d$ ~\citep{Freeman_1970}. Notice that related to this scale length, in this paper, we will use the optical radius $R_{opt}$, defined as the radius encompassing $83 \%$ of the total luminosity and proportional to the stellar disc scale length: $R_{opt}= 3.2 R_d$ (the details of this choice are expressed at length in \citealp{Persic_1996}).
In order to explain the observed rotation curves (RCs) of disc systems, it is necessary to assume the presence of a spherical DM halo surrounding 
the galaxies (~\citealp{Faber_1979, Rubin_1985} and e.g. \citealp{Salucci_2019}).

A very interesting feature of spiral galaxies is that the bigger they are, the more luminous they are and the higher rotational velocities they show.   
Moreover, when their RCs, with the radial coordinate normalised with respect to their optical radius $R_{opt}$,  are put together, they appear to follow 
a {\it universal trend} (first shown in Fig. 4 in ~\citealp{Rubin_1985}, then in \citealp{Persic_1991, Persic_1996, Rhee_1996,Roscoe_1999, Catinella_2006, Noordermeer_2007, Salucci_2007, Lopez_2018} and e.g.  \citealp{Salucci_2019}). From small to large galaxies, the RCs have higher and higher velocities and  profiles that gradually change. See also the top panel in Fig. \ref{RC_Salucci} in Appendix \ref{RC_Normal_Spirals}. 

By means of the {\it "universal rotation curve} (URC) method", which involves groupings of similar RCs and their mass modelling, it is possible to construct an analytic function that gives a good description of all the rotation curves of the local spiral galaxies within a spherical volume 
$\simeq (100 \,Mpc)^3$.
The URC method was applied for the first time in ~\cite{Persic_1991}. This was followed by a series of three works: ~\cite{Persic_1996} (Paper I), ~\cite{Salucci_2007} (Paper II) and ~\cite{Karukes_2017} (Paper III) , where the URC method gave deeper results related to {\it normal spirals}, also called  {\it high surface brightness} (HSB) spirals, and {\it dwarf disc} ({\it dd}) galaxies. A subsequent work confirmed the above results with 
up to 3100 disc galaxies and highlighted the existence of {\it tight scaling relations} among the properties of spirals with different size \citep{Lapi_2018}. 

Let us underline that the concept of {\it universality} in the RCs means that all of them  
can be described by the same analytical function as long as expressed in terms of the normalised radius and of one global parameter of the galaxies, such as magnitude, luminosity, mass or velocity at the optical radius ($V_{opt} \equiv V(R_{opt})$). 
Therefore, the {\it universal rotation curve} (URC) is the circular velocity at a certain radius $r$ given by (e.g.)
$V(r/R_{opt}, L) $, where $L$ is the galaxy's luminosity.  See the bottom panel in Fig. \ref{RC_Salucci} in Appendix \ref{RC_Normal_Spirals}.
Obviously, the URC does not change even using,
instead of $R_{opt}$, any other radial coordinate proportional to the stellar disc scale length $R_d$\footnote{The results of the paper remain unchanged for any chosen radial coordinate if expressed in units of $\lambda R_d$, with any $\lambda$ value ranging from 1 to 4.}.

The URC is a very powerful tool since, given the observation of few properties (such as $R_d$ and $L$) of a certain galaxy, it is possible to deduce its rotation curve 
and all its properties. 
\\
\\
In this paper (IV), we investigate the concept of the URC, the resulting mass models and the scaling relations in {\it Low Surface Brightness} (LSB) disc galaxies, comparing them to the results of other disc galaxies of a different Hubble type. 

LSB galaxies are rotating disc systems which emit an amount of light per area smaller than normal spirals. They are locally more isolated than other kinds of galaxies (e.g. \citealp{Bothun_1993, Rosenbaum_2004}) and  likely evolving very slowly with very low star formation rates. This is suggested by colors, metallicities, gas fractions and extensive population synthesis modelling (e.g. \citealp{vanderHulst_1993, McGaugh_1994, deBlock_1995, Bell_2000}). 
As we see in radio synthesis observations, LSB galaxies have extended gas discs with low gas surface densities and high $M_{HI}/L$ ratios (e.g. \citealp{vanderHulst_1993}), where $M_{HI}$ is the mass of the HI gaseous disc. The low metallicities make the gas cooling difficult and in turn the stars difficult to form (e.g. \citealp{McGaugh_1994}). 
LSBs are required to be dominated by DM, as shown by the analysis of their Tully-Fisher relation (e.g. \citealp{Zwaan_1995}) and of their individual rotation curves (e.g. \citealp{deBlok_2001, deBlok_2002}).

The LSB sample used in this work involves 72 galaxies selected from literature, whose optical velocities  
span from $\sim 24$ km/s to $\sim 300$ km/s, covering the values of the full population.
Our analysis of LSBs  by means of the URC method is triggered by the result shown in Fig. \ref{RCs_3D}, from which we can see that the LSBs rotation curves  gradually change very orderly from small to large galaxies (or equally from objects with small to large optical velocities $V_{opt}$).  
Following the URC method,
the sample of galaxies is divided in different velocity bins, according to their increasing values of $V_{opt}$.   
A double normalisation of all the RCs is performed with respect to: {\it i)} their own $R_{opt}$, along the radial axis, and {\it ii)} their own $V_{opt}$, along the velocity axis. In these specific coordinates, in each velocity bin, the RCs are all alike.
Then, the double normalised coadded rotation curves, a kind of average rotation curve for each velocity bin, are constructed.  
The analysis continues with their {\it mass modelling}, yielding the distribution of luminous  and dark matter in structures with different $V_{opt}$. 
This is followed by the denormalisation process, which gives the structural parameters  of  each object of the sample, and allows us to obtain the related scaling relations for the LSBs.  
The internal scatter of the  found  scaling relationships is larger (three times or more) than the analogous ones in normal spirals. 
A similar finding also emerged in the case of dwarf disc ({\it dd}) galaxies ~\citep{Karukes_2017}.
Remarkably, the scatter in the {\it dd}  relationships was reduced after the introduction of a new quantity, the {\it compactness} of the luminous matter distribution $C_*$, that indicates how the values of $R_d$ vary in galaxies with the same stellar disc mass.
Therefore, such results statistically suggest the introduction of the compactness also in the  
analogous LSBs scaling relationships. The previous steps lead to the construction of the URC for the LSBs, which is one of the main goals of this work. 
Finally, in analogy to ~\cite{Karukes_2017}, we also investigate the compactness of the dark matter distribution $C_{DM}$ and its relation  
to $C_*$.  
\begin{figure}
\begin{center} 
\includegraphics[width=0.49\textwidth,angle=0,clip=true]{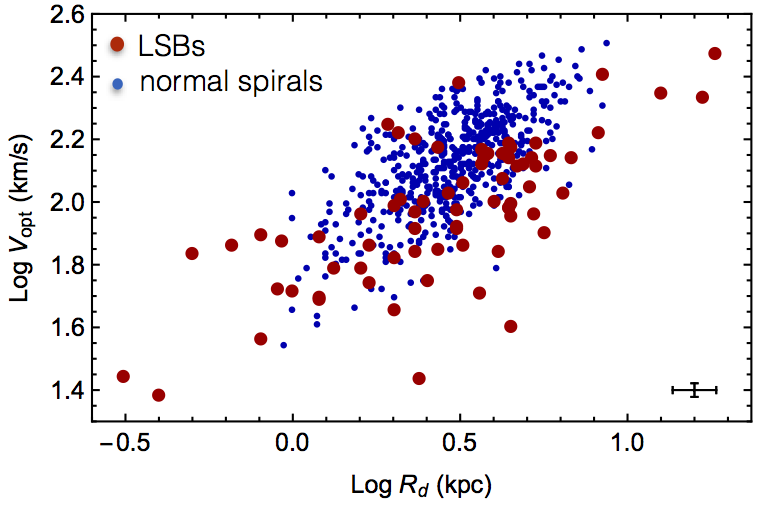} 
\caption{Optical velocity versus disc scale lengths in LSB galaxies ({\it red}) and in normal spirals ({\it blue}) ~\citep{Persic_1996}.
 The typical fractional uncertainties are 5\% in $V_{opt}$ and 15\% in $R_{d}$, as shown in the right-down corner. }
\label{Log_Vopt_vs_Ropt}
\end{center}
\end{figure} 
 \begin{figure*}
\begin{center} 
\includegraphics[width=1\textwidth,angle=0,clip=true]{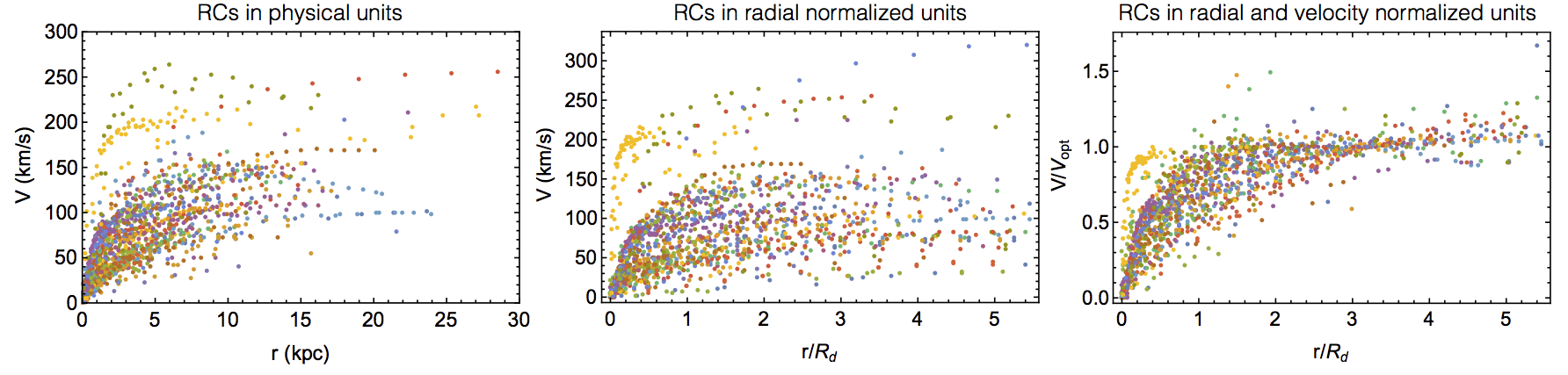}
\caption{LSBs rotation curves  (each one in different color) in physical units ({\it first panel}), in normalised radial units ({\it second panel}) 
and in double normalised radial and velocity units ({\it third panel}). See also Appendix \ref{Rotation_curves_in_physical_units}.}
\label{RC_together}
\end{center}
\end{figure*}
\begin{figure*}
\begin{center} 
\includegraphics[width=0.7\textwidth,angle=0,clip=true]{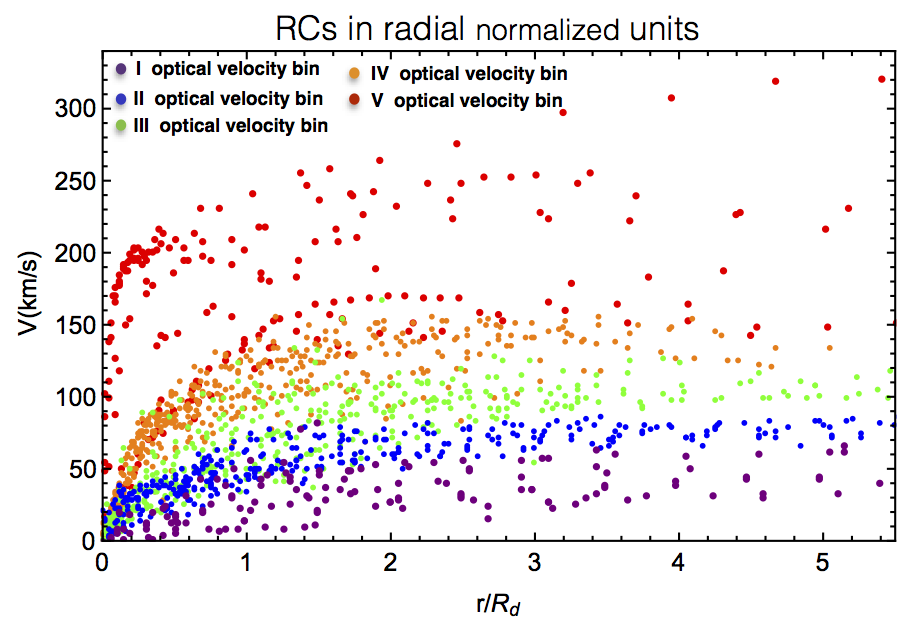}
\caption{LSBs rotation curves (in normalised radial units) grouped in five optical velocity bins. In this and in the following figures, {\it purple, blue, green, orange} and {\it red} colors are referred to the rotation curves of the I, II, III, IV and V optical velocity bins, respectively.}  
\label{InnerOuterSlope}
\end{center}
\end{figure*}
\\
\\
The structure of this paper is the following: in Section \ref{The sample}, we describe our sample of LSB galaxies; in Sections \ref{The coadded RC of LSB galaxies}-\ref{The URC of LSB galaxies}-\ref{Denormalisation of the URC mass model}, the URC method and the analysis of the LSBs structural properties are described in detail; in Section \ref{The scaling relations} we obtain the  LSBs scaling relations and we compare them to those of other disc systems; 
in Sections \ref{The compactness as the third parameter in the URC}-\ref{The correlation between $C_*$ and $C_{DM}$},
the concept of {\it compactness} is introduced and the URC-LSB is built; finally, in Section \ref{Conclusions}, we  comment on our main results.
\\
\\
The distances are evaluated from the recessional velocity assuming $H_0= 72$ km s$^{-1}$Mpc$^{-1}$.

\section{The LSB sample and the rotation curves universal trend}\label{The sample}
\noindent
We consider 72 rotating disc galaxies classified as "low surface brightness" in literature (see Tab. \ref{LSB_References_a} in Appendix \ref{LSBs sample and references}).
In the very majority of cases the authors classify a galaxy as LSB when the face-on central surface brightness $\mu_0 \gtrsim 23\,  mag \,arcsec^{-2}$ in the B band.
We select our sample according to the following criteria:
\\
{\it i}) the rotation curves extend to at least $\simeq 0.8 \,R_{opt}$ (when $V_{opt}$ is not available from observation, it can be extrapolated since 
from $\simeq1/2\, R_{opt}$ to  $2 \, R_{opt}$, the RCs are linear in radius with a small value of the slope);
\\
{\it ii}) the RCs are symmetric, smooth (e.g. without strong signs of non circular motions) and 
with an average fractional internal uncertainty lesser than $ 20 \%$. 
In short we eliminated RCs   that in no way can be mass-modelled without huge uncertainties; 
\\
{\it iii}) the galaxy disc scale length $R_d$ and the inclination function $1/sin \,i$ are known within $30 \%$ uncertainty. 

The selected 72 LSBs have optical velocities $V_{opt}$ spanning from $\sim 24$ km/s to $\sim 300$ km/s; the sample of  rotation curves consists of  
1614 independent $(r, V)$ measurements. When the RCs, expressed in normalised radial units, are put together, see Fig. \ref{RCs_3D}, they show an universal trend analogous to that of the 
the normal spirals (Fig. \ref{RC_Salucci} in Appendix \ref{RC_Normal_Spirals}).  Then, given the observed trend in LSBs and the relevance of the URC method, we search  our sample of LSBs for  a universal rotation curve and for the related scaling relations among the galaxy's structural parameters.  

In Fig. \ref{Log_Vopt_vs_Ropt}, the values of the stellar disc scale lengths $R_d$ and the optical velocities $V_{opt}$
measured in LSBs are shown and compared to those measured in normal spirals. A larger spread in the 
former case is clearly recognizable. This feature will be used later to explain the need of introducing a new structural variable,  the  {\it compactness}.

Finally, it is useful to stress that previous studies on individual LSB galaxies reveal in the mass profiles of these objects  the presence of 
an exponential stellar disc, an extended gaseous disc at very low density (e.g. \citealp{deBlok_1996}) and the presence of a spherical DM halo, likely with a  core profile (e.g. \citealp{deBlok_2001, deBlok_2002, Kuzio_de_Naray_2008}).

\section{The coadded rotation curves of LSB galaxies}\label{The coadded RC of LSB galaxies}
\noindent
The individual rotation curves (in normalised radial units) shown in Fig. \ref{RCs_3D} motivate us to proceed, also in LSB, 
with the URC method, analogously to what has been done on the high surface brightness spiral galaxies \citep{Persic_1996, Lapi_2018} and dwarf discs \citep{Karukes_2017}. It is useful to anticipate here that the average scatter of the RCs data from a fitting surface (as the URC in Fig. \ref{URCpoints}) is $\Delta V /\, V \simeq 8\%$ (taking into account the observational errors, the systematics and the small non circularities in the motion). This small value gives an idea of the universality of the LSBs rotation curves expressed in normalised radial units.
\\
\\
Among the first steps, the URC method \citep{Persic_1996} requires to make the galaxies RCs  as similar as possible (in radial extension, amplitude and profile) by introducing the normalisation of their coordinates and an eventual galaxies binning.
Let us notice that the justification for these starting steps comes from the analogous  process performed in  spirals and from a qualitative inspection of LSB RCs.  
Finally, the goodness of the results will show the goodness of the method.

The characteristics of the RCs in physical and normalised units are visible in Fig. \ref{RC_together}:  
\\
{\it i}) in the first panel, the RCs are expressed in physical units, they appear to be different in radial extension, amplitude and profile; 
\\
{\it ii}) in the second panel, the RCs are expressed in normalised radial units with respect to their disc scale length $R_d$. 
Their radial extensions are made more similar. Indeed, most of the data are extended up to $\simeq 5.5 R_d$;   
\\
{\it iii})  in the third panel, the RCs are expressed in double normalised units with respect to their disc scale length $R_d$ and optical velocity $V_{opt}$,  along the radial and the velocity axis respectively. The rotation curves in such specific units are comparable also in their amplitude.

\begin{figure*}
\begin{center} 
\includegraphics[width=1\textwidth,angle=0,clip=true]{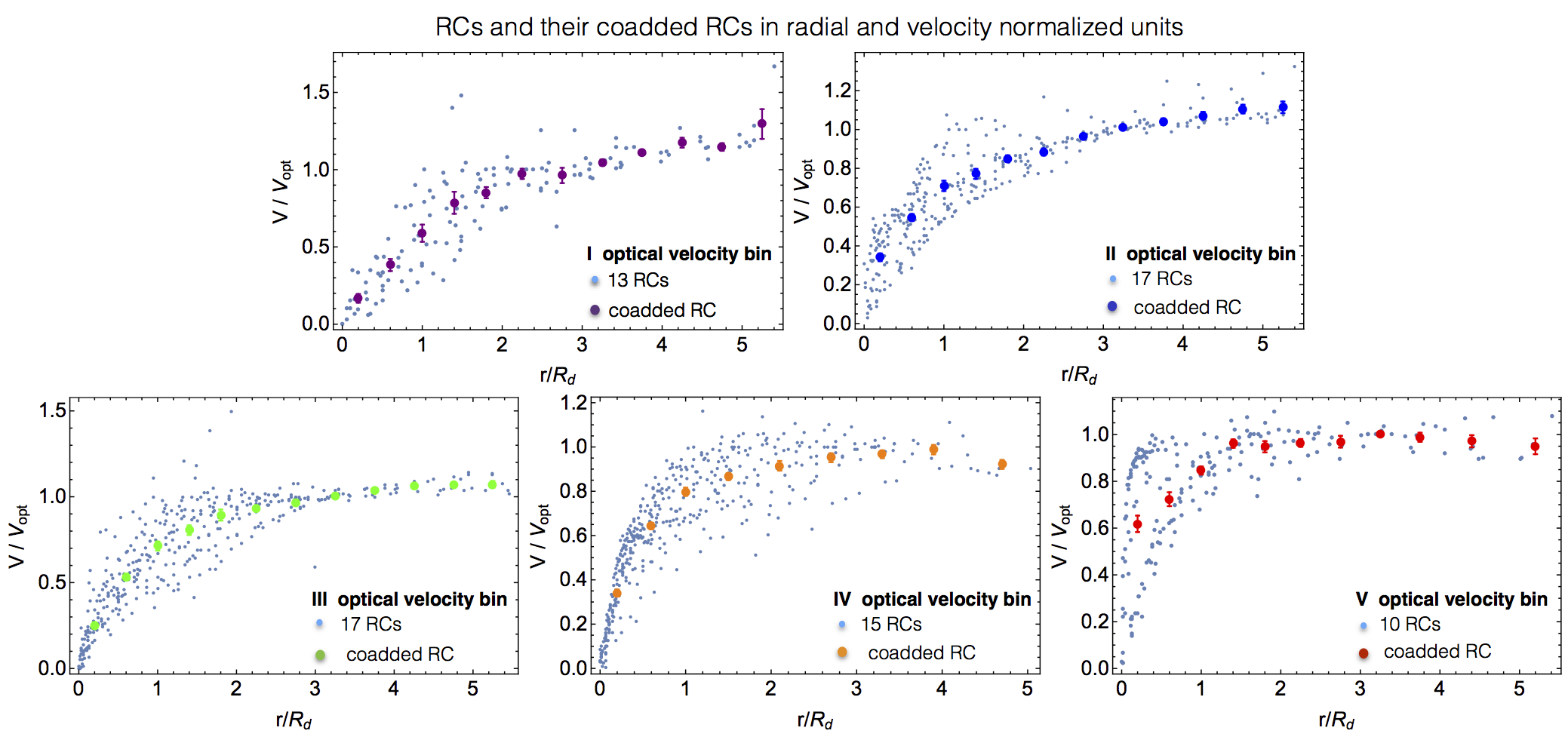}
\caption{In each of the five panels: LSBs double normalised rotation curves for each of the five optical velocity bins ({\it grey points}). Also shown the corresponding coadded rotation curves ({\it larger coloured points}) for each of these five bins. Notice that part of the scatter in the five profiles will  be eliminated by introducing the compactness in the URC. See Section \ref{The compactness as the third parameter in the URC}.}  
\label{RC_Vel_bins}
\end{center}
\end{figure*}

Overall, the double normalisation makes the 72 RCs more similar, apart from their profiles.
However,  when these RCs are arranged in 5 optical velocity bins according to their increasing $V_{opt}$ as in Fig. \ref{InnerOuterSlope}, we realise that the {\it double normalised} RCs profiles belonging to one of these bins are very similar among themselves but clearly different from those of the RCs in other optical velocity bins (see Fig. \ref{RC_Vel_bins}).    

We have chosen to build five $V_{opt}$ bins as a compromise between having a large number of data for each coadded RC and a large number of coadded RCs.
Particularly, the binning in five groups is suggested by the fact that, since the sample includes 72 objects, 10-15 galaxies are the minimum number in each optical velocity bin in order to create suitable coadded RCs (that will be described in the next paragraphs) and to eliminate statistically observational errors and small non circularities from the individual RCs.

In detail, the number of galaxies in each bin, the span in $V_{opt}$, the average optical velocity $\langle V_{opt} \rangle$, the average stellar disc scale length $\langle R_d \rangle$, the number of galaxies and of the ($r$, $V$) data are all reported in Tab. \ref{LSB_Tab2}.

We also point out  Fig. \ref{RC_Vel_bins_b} in Appendix \ref{Double_norm_RC_in_Vbin}, where the rotation curves, grouped in their velocity bins, are compared in physical and double normalised units.
%, 
\\
\\
After that all the RCs are double normalised, we perform the {\it radial binning} in each of the five {\it optical velocity bins}. 
Similarly to the velocity binning process, we have chosen 
$\simeq 11$  normalised radial bins as a compromise between having a large number of data for each radial bin and a large number of radial bins for each coadded RC. Moreover, we required that the inner radial bins (for $r\leq 2R_d$) and the outer radial bins (for $r> 2R_d$) included a minimum of 13 and 5 measurements, respectively. In detail, for the I, the II and the III 
optical velocity bins, the radial {\it normalised} coordinate is divided in 12 bins; the first 5 have a width of 0.4 and the remaining a width of 0.5. For the IV and the V velocity bins, for statistical reasons, we adopt a different division of the radial coordinate. In the IV velocity bin we adopt 3 radial bins of width 0.4, 5 of width 0.6 and the last one of width 1. In the V velocity bin, we adopt  5, 4 and 2 radial bins of widths 0.4, 0.5 and 0.8, respectively. The number of data per radial bin is reported in Tab. \ref{LSB_Tab2a} - \ref{LSB_Tab2b} in Appendix \ref{Radial_binning}. Reasonable variations of the positions and amplitudes of the radial bins do not affect the resulting coadded RCs.  

Therefore, for each of the five  $V_{opt}$ bins , in every $k$-radial bin we built  there are $N_{k}$ double normalised velocities $v_{ik}$, with $i$ running from 1 to $N_k$. Their average value is given by: $V_k= \frac{\sum_{i=1}^{N_{k}} v_{ik}}{N_k} $, as in ~\cite{Persic_1996}. Then, by repeating this for all the radial bins of each of the five $V_{opt}$ bins,  
we obtain the five double normalised {\it coadded} RCs shown in Fig.\ref{RC_Vel_bins}.
The {\it standard error of the mean} we  consider in this  work, is 
\begin{eqnarray}
\label{Errors_coadded_RC}
\delta V_k =  \sqrt{\frac{\sum_{i=1}^{N_k} (v_{ik} -V_k)^2 }{N_k(N_k-1)}}  \quad .
\end{eqnarray}   
\\
\\
In short, the above coadded RCs can be considered as the  average rotation curves of galaxies of similar properties as, e.g.,  $V_{opt}$. 
It is worth emphasizing the advantages of these RCs: their building erases the peculiarities and much reduces the observational errors  of the individual RCs. This yields to a  universal description of the kinematics of LSBs by means of 5 extended and smooth  RCs whose values have an uncertainty at the level  of $5\%-15\%$. In Fig. \ref{Binning_Double_Norm} the five coadded RCs are shown together:
\begin{figure*}
\begin{center} 
\includegraphics[width=1\textwidth,angle=0,clip=true]{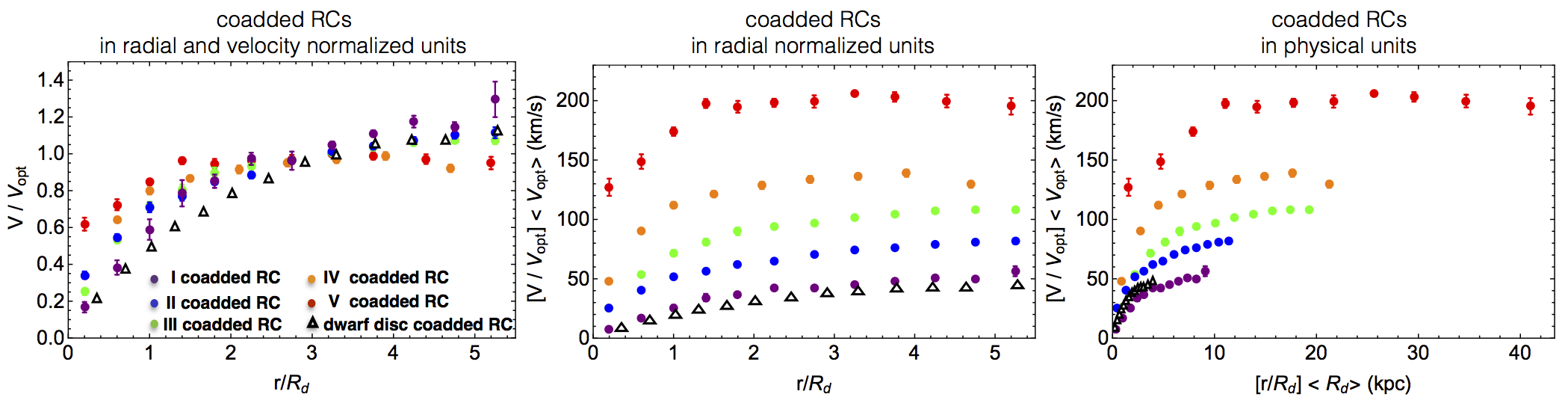}
\caption{Coadded rotation curves for the five velocity bins in double normalised units ({\it first panel}), in physical velocity units ({\it second panel}) and in physical units along both the velocity and radial axes ({\it third panel}). The black empty triangles are the coadded RC for the dwarf disc galaxies ~\citep{Karukes_2017}.}  
\label{Binning_Double_Norm}
\end{center}
\end{figure*}
\begin{table}
\begin{tabular}{p{1.1cm}p{1.4cm}p{1.2cm}p{0.8cm}p{0.8cm}p{0.7cm}}
\hline
$V_{opt}$ bin  &    $V_{opt}$ range     &     N.galaxies     &    $\langle V_{opt} \rangle$   &    $\langle R_d \rangle$       &    N.data     \\
               &      km/s            &                             &                        km/s                     &              kpc             &                   \\
  (1)          &       (2)                 &          (3)               &                              (4)                   &                  (5)            &        (6)              \\
\hline
1              &   24-60            &     13                  &                        43.5                     &              1.7                      &         151            \\
2              &   60-85            &     17                  &                         73.3                     &              2.2                      &         393           \\
3              &   85-120         &      17                  &                        100.6                   &               3.7                      &          419           \\
4              &   120-154       &      15                 &                        140.6                   &               4.5                      &          441           \\
5              &   154-300       &      10                   &                         205.6                   &              7.9                     &          210          \\
\hline 
\end{tabular}
\caption{LSB velocity bins. Columns: (1) i - velocity bin; (2) range values for $V_{opt}$; (3) number of LSB galaxies in each velocity bin; (4) average value of $V_{opt}$ evaluated from the individual galaxies; (5) average value of $R_d$ evaluated from the individual galaxies; (6) number of ($r$, $V$) data from the individual galaxies.}
\label{LSB_Tab2}
\end{table}
\\
{\it i}) in the first panel, they are expressed in double normalised units covering a very small region
in the ($V/V_{opt},R/R_{opt}$) plane ;
\\
{\it ii}) in the second panel, they are expressed in physical velocity units. These coadded RCs are obtained by multiplying the previous coadded RCs  by the corresponding $\langle V_{opt} \rangle$ (reported in Tab. \ref{LSB_Tab2}). 
\\
{\it iii}) in the third panel, the coadded RCs are expressed in physical units both along the velocity and the radial axes. 
They are obtained by multiplying the previous coadded RCs by the corresponding $\langle R_d \rangle$ reported in Tab. \ref{LSB_Tab2}.   

In Fig. \ref{Binning_Double_Norm} the difference in the profiles corresponding to galaxies with different optical velocities is evident. 
\footnote{This is explained by the very different luminous and dark mass distributions in LSBs of different sizes and optical velocities, as shown in the next section.} 

All the data shown in Fig. \ref{Binning_Double_Norm} can be recast by means of Tab. \ref{LSB_Tab2a} - \ref{LSB_Tab2b} (in Appendix \ref{Radial_binning}) and Tab. \ref{LSB_Tab2}.

\section{The mass modelling of the coadded rotation curves}\label{The URC of LSB galaxies}
\noindent
In this section we investigate the coadded rotation curves, normalised along the radial axis (see second panel in Fig. \ref{Binning_Double_Norm}), whose data are listed in Tab. \ref{LSB_Tab2a} - \ref{LSB_Tab2b} in Appendix \ref{Radial_binning}. We model the coadded RCs data, as in normal spirals \citep{Salucci_2007}, with an analytic function $V(r)$  which includes the contributions from the stellar disc $V_d$ and from the DM halo $V_h$:

\begin{table*}
\begin{tabular}{p{1.9cm}p{1.9cm}p{1.9cm}p{1.9cm}p{2.3cm}p{1.9cm}p{1.6cm}p{1.2cm}}
\hline
Vel. Bin  &   $ \langle V_{opt} \rangle $    & $ \rho _0 $        &   $  R_c $        &    $  M_d  $          &    $  M_{vir}  $  &      $\alpha (R_{opt})$            &        $k$         \\
                &    $km/s$     & $10^{-3} \, M_{\odot} /pc^3$     &             kpc                         &      $10^{11}\,M_{\odot}$                &     $10^{11}\,M_{\odot}$      &                                         &                                \\
  (1)          &                (2)                               &              (3)                             &                         (4)                      &       (5)                                  &                    (6)               &        (7)         &        (8)   \\
\hline
1                &    43.5    &   $3.7 \pm 1.4$                          &     $10.7 \pm 4.3 $                  &              $(8.8 \pm 1.8)\times 10^{-3} $               &       $ 1.0 \pm 0.4 $                                &            0.37                    &       0.36      \\
2              &   73.3   &   $5.1 \pm 1.1$                          &     $12.8 \pm 3.0 $                  &             $(3.8 \pm 0.3)\times 10^{-2}$                   &          $2.4  \pm 0.9$                               &            0.49                    &       0.44       \\
3              &          100.6      &        $3.7 \pm 0.5$                          &     $17.1 \pm 1.9 $                  &             $(13.0 \pm 0.5)\times 10^{-2}$                 &        $   4.0 \pm 1.3 $                               &            0.52                    &        0.47      \\
4              &      140.6       &   $1.7^{+1.8}_{-1.1}$                 &     $30^{+40}_{-22}$       &             $(4.2 \pm0.4)\times 10^{-1}$                &         $  8.4 \pm 3.5 $                               &            0.76                   &        0.63        \\
5             &         205.6           &   $0.8^{+0.7}_{-0.4} $                &     $99^{+213}_{-87} $      &           $1.7 \pm0.1$            &         $  112  \pm 55 $                             &            0.82                   &        0.70       \\
\hline  
\end{tabular}
\caption{Relevant parameters of  the five coadded RCs. Columns: (1) i - velocity bin; (2) average value of $V_{opt}$; best fit value of (3) $\rho _0$; (4)  $R_c$; 
(5)$M_d$; (6) estimated halo virial mass according to Eq. \ref{Virial_mass}; (7) fraction of baryonic component at $ R_{opt}$
(Eq. \ref{Alpha_eq}); (8) $k$ values defined according to Eq. \ref{Denormalisation_1}.}
\label{LSB_Tab3}
\end{table*}
\begin{figure*}
\begin{center} 
\includegraphics[width=1\textwidth,angle=0,clip=true]{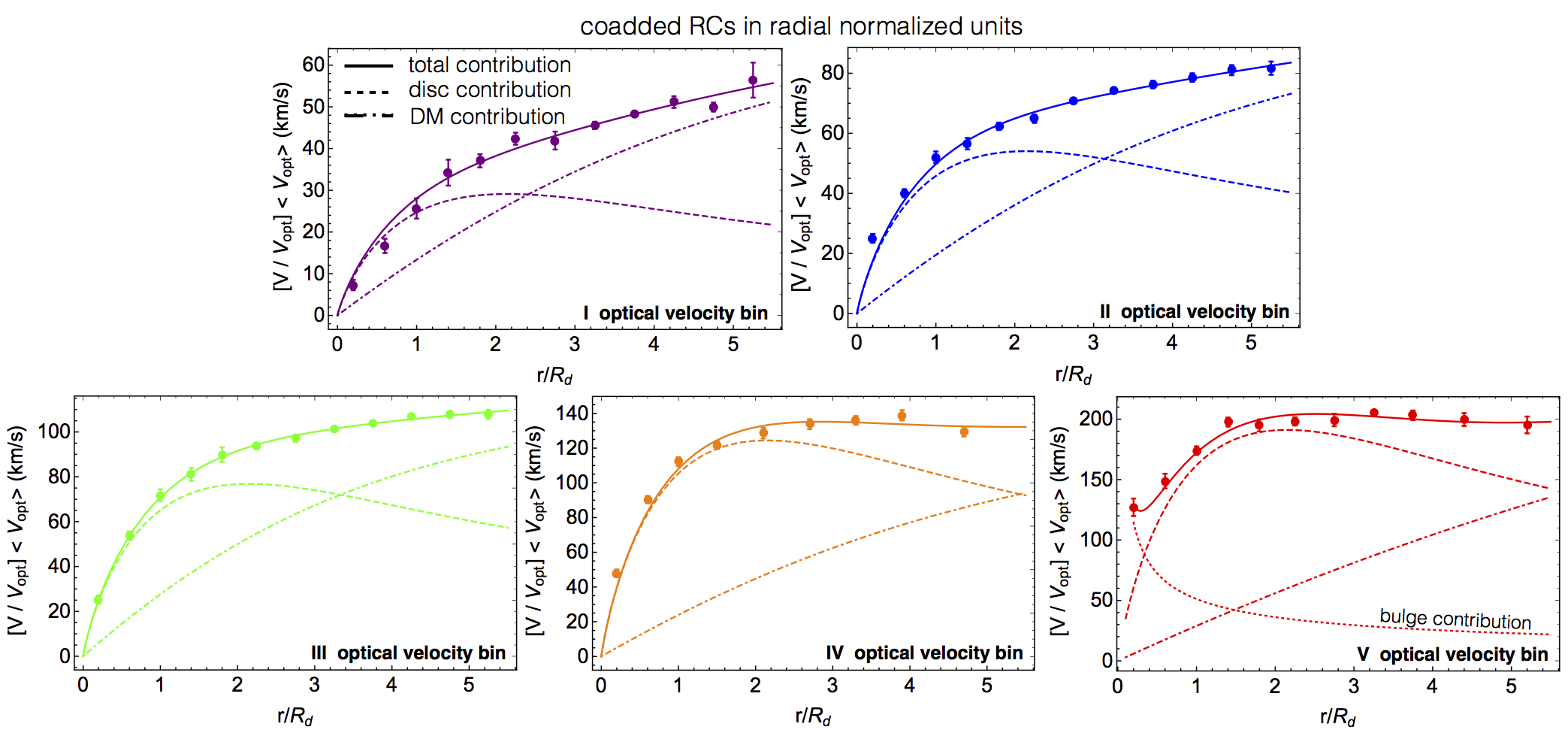}
\caption{In each of the five panels  the velocity  best-fit models to the corresponding  coadded RCs are shown. The {\it dashed, dot-dashed, dotted} and {\it solid} lines indicate the stellar disc, the DM halo, the stellar bulge and the model contribution to the circular velocities.}
\label{Velocity_1}
\end{center}
\end{figure*}
\begin{eqnarray}
\label{V_Model}
V^2(r) = V_d^2(r) + V_h^2 (r)  \quad .
\end{eqnarray}   
Let us stress that in first approximation  the inclusion in the model of a HI gaseous disc component can be neglected. In fact, the gas contribution is usually a minor component to the circular velocities, since the inner regions of galaxies are dominated by the stellar component and in the external regions, where the gas component overcomes the stellar one, the DM contribution is largely the most important ~\citep{Evoli_2011}. A direct test in Appendix \ref{Gas_component} shows that our assumption does not  affect the mass modelling obtained in this paper.
\\
\\
We describe the stellar and the dark matter component. The first one is given by the well-known Freeman disc ~\citep{Freeman_1970}, whose surface density profile is 
\begin{eqnarray}
\label{Surface_stellar_density}
\Sigma _d (r) = \frac{M_d}{2\pi R_d^2}\,  exp(-r/R_d)   \quad ,
\end{eqnarray}   
where $M_d$ is the disc mass.
Eq. \ref{Surface_stellar_density} leads to \citep{Freeman_1970}:
\begin{eqnarray}
\label{V_stellar}
V_d^2(r) = \frac{1}{2} \frac{G\, M_d}{R_d} \left(\frac{r}{R_d}\right)^2 (I_0 K_0 - I_1 K_1)      \quad ,
\end{eqnarray}   
where $I_n$ and $K_n$ are the modified Bessel functions computed at $1.6 \,x$, with $x = r/R_{opt}$.

 Finally, for the fifth optical velocity bin we will introduce a bulge component.   ~\citep{Das_2013}.
\\
\\
Concerning the dark matter component, the presence of cored profiles in LSBs is well known from individual rotation curves (see e.g. \citealp{deBlok_2001, deBlok_2002, Kuzio_de_Naray_2008}, \citealp{Bullock_2017} ). In this paper, we model the DM halo profile 
by means of the {\it cored Burkert profile} \citep{Burkert_1995, Salucci_2000a}. This halo profile has an excellent record in fitting the actual  DM halos around  disc systems of any luminosity and Hubble Types (see  \citealp{Salucci_2019};  \citealp{Lapi_2018}, \citealp{Memola_2011}, \citealp{Salucci_2012}). In addition, the Burkert profile is in agreement with weak lensing data at virial distances ~\citep{Donato_2009}.

It is however  worth to noticing that there is no sensible difference, in the mass modelling inside $R_{opt}$,  in adopting different cored DM density profiles \citep{Gentile_2004}. 
Then, we adopt the following density profile \citep{Burkert_1995}:
\begin{eqnarray}
\label{DM_density}
\rho _{DM} (r) = \frac{\rho _0 R_c ^3 }{(r+R_c)(r^2+R_c^2 )}     \quad ,
\end{eqnarray}   
where $\rho _0$ is the central mass density and $R_c$ is the core radius. Its mass distribution is:
\begin{eqnarray}
\label{DM_Mass}
M_{DM} (r)  & = &     \int_0^r 4\pi {\tilde r}^2 \rho _{DM} ({\tilde r}) \; d  {\tilde r}   =   \\ \nonumber
                    &  =  & 2 \pi \rho _0 R_c ^3 \; [ln (1+r/R_c )    \\ \nonumber
                      &  & - tg ^{-1} ( r/R_c )  +     0.5\; ln (1+( r/R_c) ^2  )]        \quad .                 
\end{eqnarray}   
The contribution to the total circular velocity is given by:
\begin{eqnarray}
\label{V_dark_matter}
V_h^2(r) = G \frac{M_{DM}(r)}{r}      \quad .
\end{eqnarray}   
\\
\\
We fit the  five coadded RCs by means of the URC model described above, which, for each coadded RC, is characterized by three free parameters, $M_d$, $\rho_0$ and $R_c$, all set to be larger than zero. Other limits for the priors of the fitting  arise from the amplitude and the profile of the coadded RCs themselves. We require that: $10^6 M_{\odot} \lesssim M_d  \lesssim  10^{12} \, M_{\odot}$ from the galaxies luminosities, $ R_c \lesssim 200  \ \frac{R_{opt}}{30 kpc} \, kpc$ to avoid solid body rotation curves in all objects and $10^{-26} \lesssim  \rho_0 \lesssim 10^{-22}\,  g/cm^3$ (the lower  limit guarantees that the dark component is able to  fit the RC allied with the luminous component, the upper limit is to make the dark matter contribution  important but not larger  than the RCs amplitudes).   Notice that these limits well agree  with  the outcomes of the  modelling of individual  RCs  as found in literature. 

The resulting best fit values for  the three free parameters ($M_d$, $\rho_0$, $R_c$) are reported in Tab. \ref{LSB_Tab3} and the best fit velocity models  are plotted alongside the coadded RCs in Fig. \ref{Velocity_1}.

In the case of the V velocity bin, we introduce a central bulge (whose presence is typical in the largest galaxies) \citep{Das_2013}.
We adopt for  the bulge velocity component the simple functional form:
\begin{eqnarray}
\label{Bulge_velocity}
V^2_{b}(r)= \alpha _b V^2_{in} \left(\frac{r}{r_{in}} \right)^{-1}  \quad ,
\end{eqnarray}   
where $\;V_{in}= 127 \,km/s\,$ and $\;r_{in}= 0.2\, \langle R_d\rangle \simeq 1.6 \, kpc$ are the values of the first velocity  point of the V coadded RC. Since $r_{in}$ is  larger than the edge of the bulge, we consider the latter as a point mass.  $\alpha _{b}$ is a parameter which  can vary from $0.2$ to $1$ 
(e.g. see  ~\citealp{Yegorova_2007}). By fitting the V coadded RC we  found: $\alpha_b = 0.8 $;  
the other best fit parameters $M_d$, $\rho_0$, $R_c$ are reported in Tab. \ref{LSB_Tab3}.
\\
\\
In Fig. \ref{Velocity_1} we realise that, in the inner regions of the LSB galaxies, the stellar component (dashed line) is dominant; while, on the contrary, in the external regions, the DM component (dot-dashed) is the  dominant one. Moreover, the transition radius\footnote{The {\it transition radius} is the radius where the DM component, dot-dashed line, overcomes the luminous component, dashed line.} between the region dominated by the baryonic matter and the region 
dominated by the dark matter increases with normalised radius when we move
from galaxies with the lowest $V_{opt}$ to galaxies with the highest $V_{opt}$. A similar behaviour was also observed in normal spiral galaxies ~\citep{Persic_1996, Lapi_2018}.

\section{Denormalisation of the coadded rotation curves}\label{Denormalisation of the URC mass model}
\begin{figure}
\begin{center} 
\includegraphics[width=0.49\textwidth,angle=0,clip=true]{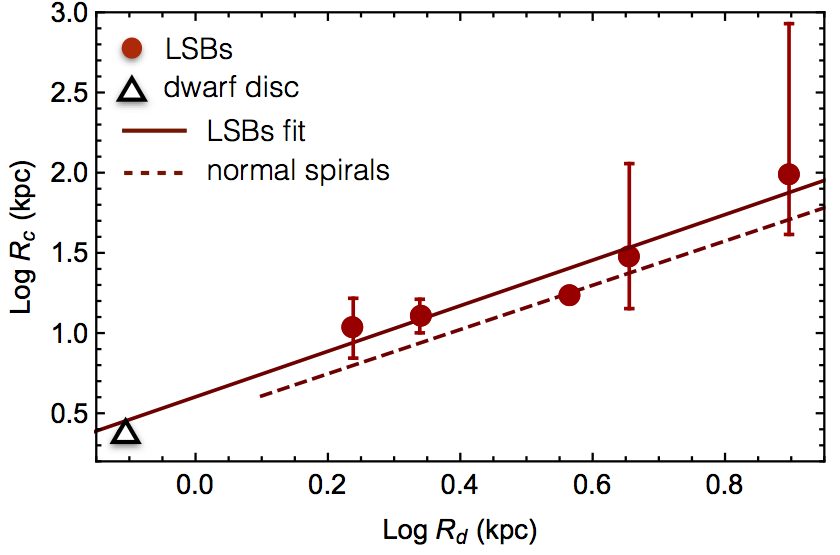}
\caption{Relationship between the DM halo core radius and the stellar disc scale length ({\it points}) and its best fit ({\it solid line}) compared to that of the normal spirals({\it dashed line}) (e.g. \citealp{Lapi_2018}). The black empty triangle represents the  relationship in dwarf disc galaxies ~\citep{Karukes_2017}.}  
\label{RC_RD_relation}
\end{center}
\end{figure}

\noindent
The mass models found in the previous section provided us with the structural parameters of the five coadded RCs. Now, we retrieve  
the properties from the individual RCs by means of the {\it denormalisation} method.
It relies on the facts that, in each velocity bin, {\it i)} all the double normalised RCs are similar to their coadded double normalised RC (see Fig. \ref{RC_Vel_bins})
and that {\it ii)} we have performed extremely good fits of the coadded RCs (see Fig. \ref{Velocity_1}). 
Thus, the relations existing for the coadded RCs are assumed to approximately hold also for the individual RCs that form each of the 5 coadded ones.  
\\ 
\\
The first relation that we apply in the denormalisation process is shown in Fig. \ref{RC_RD_relation}; 
the stellar disc scale length and the DM core radius  of the five velocity  models are strongly correlated. 
The best linear fit in logarithmic scale is:
\begin{eqnarray}
\label{Denormalisation_2}
Log \, R_c = 0.60 + 1.42\, Log \, R_d    \quad ,    
\end{eqnarray}  
The errors in the fitting parameter are shown in Tab. \ref{Error_Scaling_Relations_Tab} in Appendix \ref{Error_Scaling_Relations}. %
The relation expressed by Eq. \ref{Denormalisation_2} means that, in each galaxy  we can evaluate $R_c$ from its measured  $R_d$.
It is worth noting that a similar relation exists also in normal spirals (Fig. \ref{RC_RD_relation}).
\\
\\
\begin{figure}
\begin{center} 
\includegraphics[width=0.49\textwidth,angle=0,clip=true]{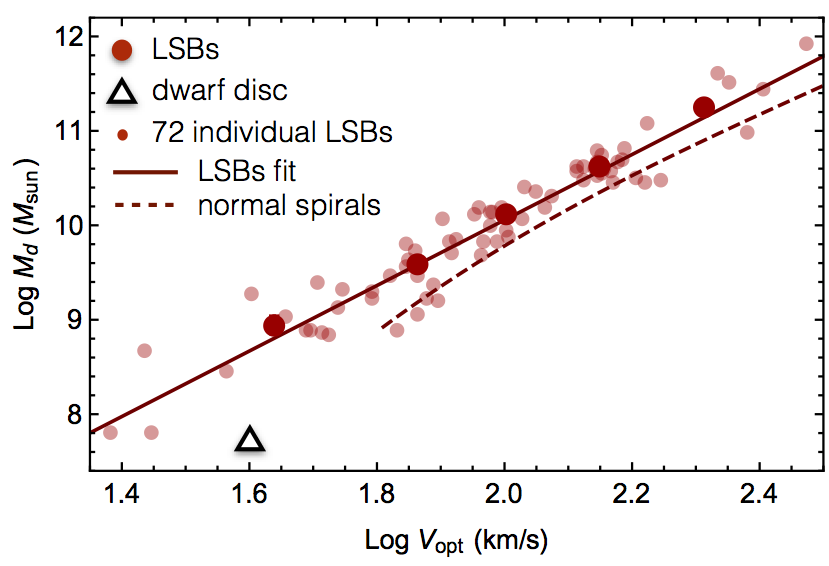} 
\caption{Relationship between the stellar disc mass and the optical velocity. The {\it large points} refer to the values of the  five velocity bins, while  
the {\it small points} refer to the values of each  LSB galaxy. The {\it solid} and the {\it dashed} lines are the best fit for LSBs and normal spirals (e.g. \citealp{Lapi_2018}). The black  triangle represents  the dwarf discs ~\citep{Karukes_2017}.}
\label{Mstar_vs_Vopt}
\end{center}
\end{figure} 
 \begin{figure*}
\begin{center} 
\includegraphics[width=0.49\textwidth,angle=0,clip=true]{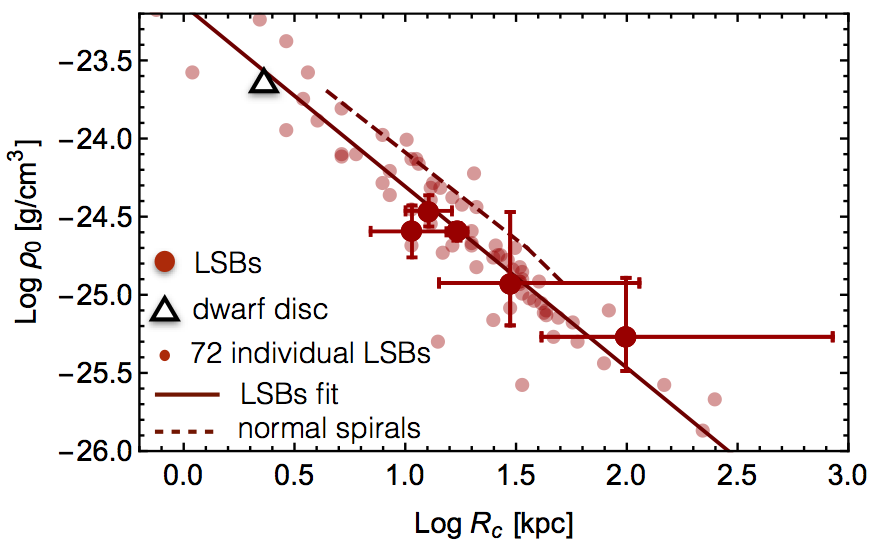} 
\includegraphics[width=0.48\textwidth,angle=0,clip=true]{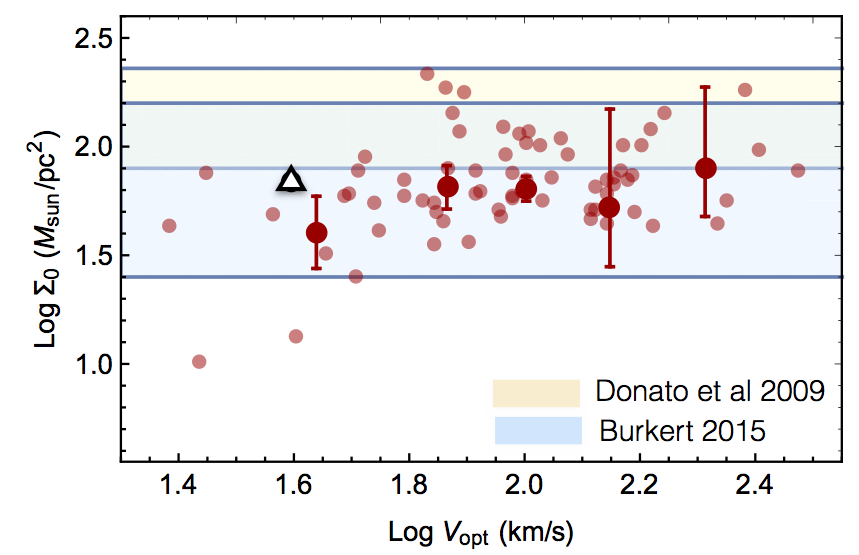} 
\caption{{\it Left} panel: the relationship between the central DM halo mass density and its core radius. 
{\it Right} panel: surface density $\Sigma _0 = \rho _0 R_c$  versus their optical velocities $V_{opt}$ (LSBs in {\it red} points). Also shown the scaling relation obtained by ~\citealp{Donato_2009} ({\it yellow} shadowed area)  and ~\citealp{Burkert_2015}  ({\it light blue} shadowed area). The black empty triangle represents the dwarf discs  ~\citep{Karukes_2017}.}
\label{Rho0Sigma0}
\end{center}
\end{figure*}
The second relation we use for the denormalisation assumes that for galaxies belonging to each $V_{opt}$ bin: 
\begin{eqnarray}
\label{Denormalisation_1}
\frac{G \, M_d}{V_{opt}^2 R_{opt}} =  k      \quad ,
\end{eqnarray}  
where the $k$ values are reported in Tab. \ref{LSB_Tab3}. $R_{opt}$ and $V_{opt}$ are measured for all the galaxies, thus
Eq. \ref{Denormalisation_1} allows us to evaluate the stellar disc mass $M_d$ for each of them.
\\
\\
As the  third step in the denormalisation process we evaluate at $R_{opt}$, for each of the  five coadded RCs, the fraction of the baryonic matter:
\begin{eqnarray}
\alpha(R_{opt}) = \frac{ V_d^2 (R_{opt})}{ V^2 (R_{opt})}   \quad .
\label{Alpha_eq}
\end{eqnarray}   
The $\alpha(R_{opt})$ values are reported in Tab. \ref{LSB_Tab3}; we assume that all the galaxies included in each optical velocity bin take the 
same value for $\alpha(R_{opt})$. Then, for  each galaxy, we write  the DM mass inside the optical radius   as:
\begin{eqnarray}
\label{Denormalisation_3}
M_{DM} (R_{opt})= [1- \alpha  (R_{opt})] V_{opt}^2 R_{opt} G^{-1}  \quad .
\end{eqnarray}   
Finally, by considering Eq. \ref{DM_Mass}-\ref{Alpha_eq}-\ref{Denormalisation_3} together with the result from the first denormalisation step, 
we evaluate the central density of the DM halo $\rho_0$ for each galaxy.
\\
\\
The structural parameters of the dark and luminous matter of the  galaxies of our sample, inferred by the denormalisation procedure, are reported in Tab. \ref{LSB_Tab4}-\ref{LSB_Tab5} in Appendix \ref{Structural_properties}. Moreover, we have  the basis to infer other relevant quantities of the galaxies structure that will be involved, in the next section, in building the scaling relations. The virial mass $M_{vir}$, that practically encloses the whole mass of a galaxy, is evaluated 
according to:
\begin{eqnarray}
\label{Virial_mass}
M_{vir}= \frac{4}{3} \, \pi \,100 \, \, \rho_{crit} \, R_{vir}^3  \quad ,
\end{eqnarray}   
where $R_{vir}$ is the virial radius and $\rho_{crit} = 9.3 \times 10^{-30} g/cm ^3$ is the critical density of the Universe.
The DM central surface density $\Sigma_0$ is evaluated by the product $\rho_0$ and $R_c$. 
The $M_{vir}$ and $\Sigma_0$ values for the objects in our sample are shown in Tab. \ref{LSB_Tab4}-\ref{LSB_Tab5} in Appendix \ref{Structural_properties}.  

\section{The scaling relations}\label{The scaling relations}

\noindent
In this section, we work out the scaling relations among the structural properties of dark and luminous matter in each LSB  galaxy. 
Let us stress that for many of the scaling relations we have no a priori insight of how they should be; in this case, the 
goal is to find a statistically relevant relationship. Then we fit the observational data  with the simple power-law model.
The errors on the fitting parameters of the various scaling relations and their standard scatters  are reported in Tab. \ref{Error_Scaling_Relations_Tab} in Appendix \ref{Error_Scaling_Relations}. Hereafter, the masses are expressed in $M_{\odot}$, the radial scale length in $kpc$, the velocities in $km/s$ and the mass densities in $g/cm^3$.
\\
\\ 
We start with the relation between the stellar disc mass and the optical velocity. Fig. \ref{Mstar_vs_Vopt} shows that the LSB data are well fitted by: 
\begin{eqnarray}
\label{MD_Vopt_fit}
Log \, M_d = 3.12 + 3.47 \, Log \, V_{opt} \quad .
\end{eqnarray}    
This relation holding for the LSBs is 
similar but not identical to the normal spirals' one. See the comparison with \cite{Lapi_2018} in Fig. \ref{Mstar_vs_Vopt}.

Next, in Fig. \ref{Rho0Sigma0} (left panel) we show the relation between the DM halo central density and the core radius, which indicates that the highest mass densities are in the smallest galaxies, as also found in normal spirals \citep{Salucci_2007}. We find:
\begin{eqnarray}
\label{rho0_Rc_fit}
Log \, \rho _0 = -23.15 - 1.16 \, Log\, R_c  \quad .
\end{eqnarray}

Moreover, we find that the central surface density follows the relationship (see Fig. \ref{Rho0Sigma0} (right panel)): 
\begin{eqnarray} 
\label{Sigma_fit}
Log \, \Sigma _0 = Log  \, (\rho _0 R_c) \simeq 1.9  \quad ,
\end{eqnarray}    
 $\Sigma _0$ is  expressed in units of $M_{\odot}/ pc^2$.
 
 Remarkably, this relationship extends itself over 18 blue magnitudes and in objects spanning from dwarf  to giant galaxies ~\citep{Spano_2008, Gentile_2009, Donato_2009, Plana_2010, Salucci_2012, McGaugh_2019, Chan_2019}.  
\begin{figure}
\begin{center} 
\includegraphics[width=0.48\textwidth,angle=0,clip=true]{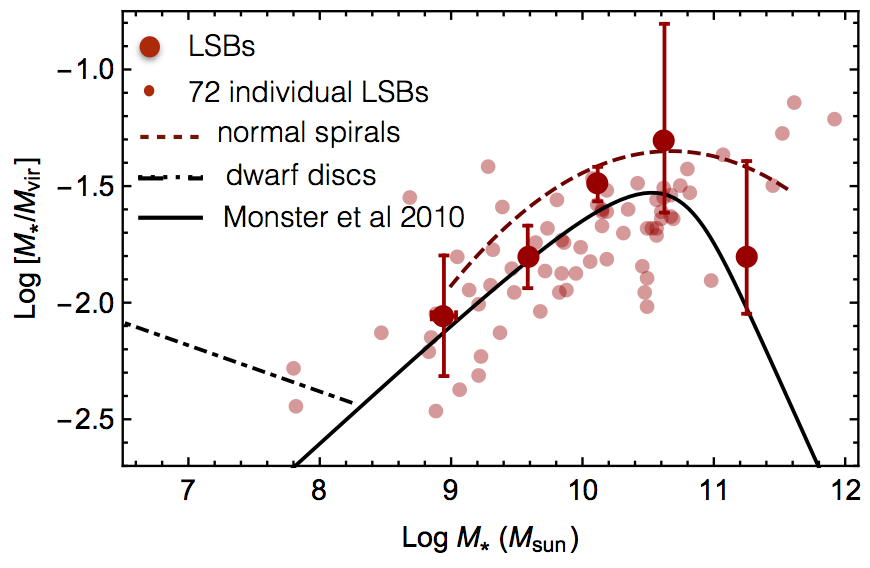}
\caption{Fraction of baryonic matter in LSBs versus their  mass in stars ({\it points}) compared with that  of normal spirals ({\it dashed} line) \citep{Lapi_2018}, of other Hubble Types ({\it black solid} line)  \citep{Monster_2010} and of dwarf discs ({\it black dot-dashed} line) \citep{Karukes_2017}.}
\label{MdiscMvirMdisc}
\end{center}
\end{figure}
\begin{figure*}
\begin{center} 
\includegraphics[width=1\textwidth,angle=0,clip=true]{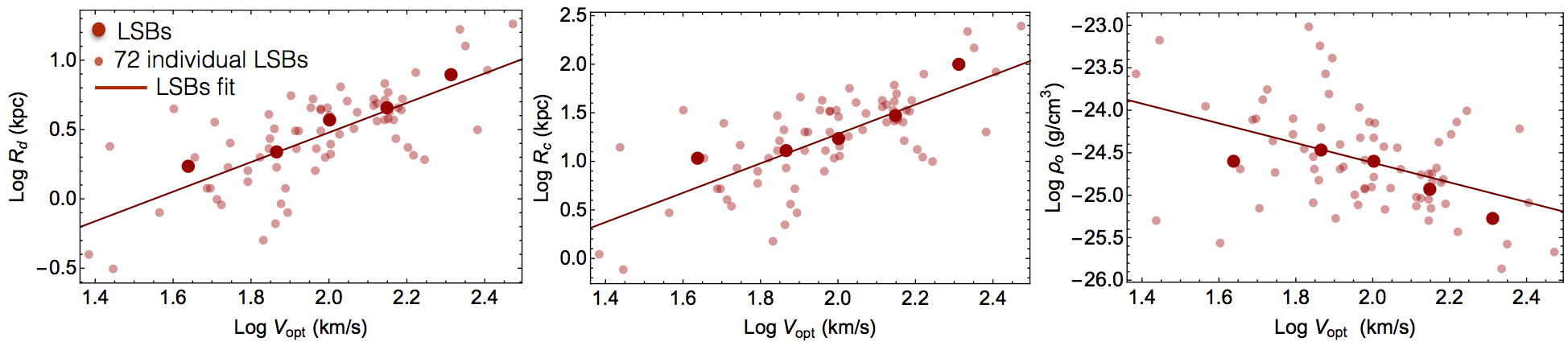}
\caption{LSBs relationships between {\it a)} the stellar disc scale length, {\it b)} the DM core radius and {\it c)} the central DM core density versus the optical velocity ({\it first, second} and {\it third} panel).}
\label{MD_Mvir}
\end{center}
\end{figure*}

Then, we consider the baryonic fraction (complementary to the DM fraction) relative to the entire galaxies, namely, the ratio between the stellar mass $M_*\equiv M_d$ in LSBs and the virial mass $M_{vir}$, that practically represents  the whole dark  mass of a galaxy. Fig. \ref{MdiscMvirMdisc} shows that the lowest fraction of baryonic content is in the smallest galaxies (with the smallest stellar disc mass $M_d$). We note that this ratio  increases going towards larger galaxies and then reaches a plateau from which it decreases for  the largest galaxies.    
This finding is in agreement with the inverse "U-shape" of previous works relative to  normal spirals ~\citep{Lapi_2018}. 
Furthermore, our result seems to follow a trend similar to that found in ~\cite{Monster_2010}, concerning all Hubble Types \footnote{In ~\cite{Monster_2010}, the stellar mass $M_*$ can indicate 
the mass enclosed in a disc and/or in a bulge).}. 
The result points to a less efficient star formation in the smallest LSBs.
\\
\\
Finally, we work out  the relationships needed to establish  $V_{URC}(R;R_{opt},V_{opt})$, the URC-LSB in physical units (as in \citealp{Persic_1996}).
Straightforwardly, we are looking for the universal function $V_{URC}(r/R_{opt}, V_{opt})$\footnote{Hereafter, we express the normalised radial coordinate in terms of the optical radius $R_{opt}$, instead of $R_d$, in order to facilitate the comparison  with previous works on the URC.}, able to reproduce analytically the LSBs RCs in Fig. \ref{RCs_3D}. 
  
  This implies that $M_d$, $R_d$, $R_c$ and $\rho_0$ have to be expressed as a function of $V_{opt}$. Thus, we use  Eq. \ref{MD_Vopt_fit} and  the following relations, obtained after the denormalisation process:
\begin{eqnarray}
\label{Rd_Vopt_fit}
Log \,R_d =  -1.65 + 1.07 \, Log \, V_{opt} \\  \nonumber
Log \,R_c =  -1.75 + 1.51 \, Log \, V_{opt} \\ \nonumber
Log \, \rho_0 =  -22.30 - 1.16 \, Log \, V_{opt}   
\end{eqnarray}   
See Fig. \ref{MD_Mvir}.
We note that the above relations (Eq. \ref{MD_Vopt_fit}-\ref{Rd_Vopt_fit}) show a large scatter, on average $\sigma \simeq 0.34$ dex, more than three times the value ($\sigma \simeq 0.1$ dex  in \citealp{Lapi_2018}, \citealp{Yegorova_2007}) found in normal spiral galaxies for the respective relations. This poses an issue to the standard procedure \citep{Persic_1996} to build the URC in physical units.  

In the  previous sections we have found a universal function to reproduce the double normalised rotation curve of LSBs 
$V(r/R_{opt})/V(R_{opt})$. Now we are looking for a universal function to reproduce the RC in physical units $ V(r)$. In spiral galaxies this is simple since $M_d$, $R_d$, $R_c$ and $\rho_0$ are closely connected.

\section{The compactness as the third parameter in the URC}\label{The compactness as the third parameter in the URC}
\begin{figure}
\begin{center}
\includegraphics[width=0.46\textwidth,angle=0,clip=true]{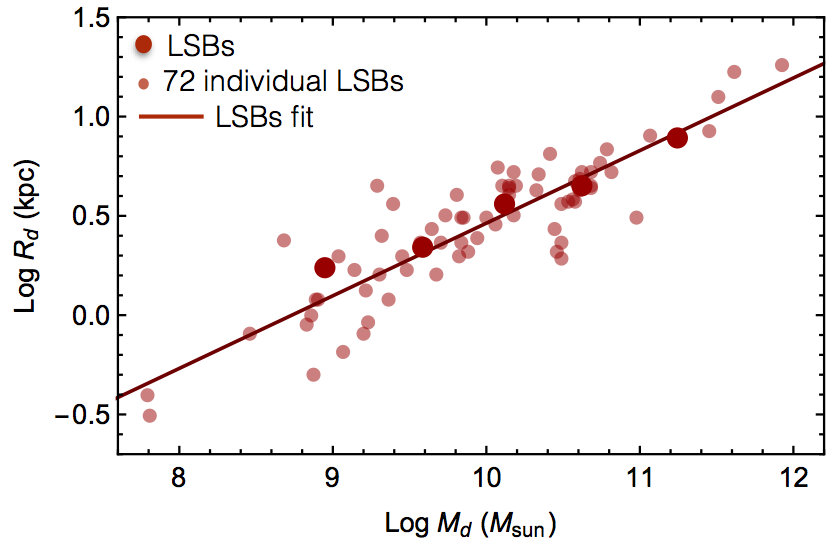} 
\caption{Relationship between the stellar disc scale length and the stellar disc mass.}
\label{RD_MD}
\end{center}
\end{figure}

 \noindent
 We can reduce the scatter in the LSBs scaling relations and proceed with the URC building by introducing a new parameter, the {\it compactness} 
of the stellar mass distribution $C_*$. This parameter was first put forward  in \cite{Karukes_2017} to cope with a similar large scatter in the above scaling relations of the {\it dd} galaxies.
In short the large scatter in the previous relationships  is due to the fact that galaxies with the same stellar disc mass $M_d$ (or $V_{opt}$) can have a very different size for  $R_d$ (i.e. $Log \, R_d$ can vary almost 1 dex). We define this effect with  the fact that LSBs have a different "stellar compactness" $C_{*}$; see Fig. \ref{Log_Vopt_vs_Ropt} and Fig. \ref{RD_MD}. 

We define $C_*$,   starting from the  best fit linear relation (see Fig. \ref{RD_MD}):
\begin{eqnarray}
\label{Stellar_Compactness_1}
Log \, R_d = -3.19 + 0.36 \,Log \, M_d \quad 
\end{eqnarray}
and,  according to ~\cite{Karukes_2017}, we set the stellar compactness through the following relation:
\begin{eqnarray}
\label{Stellar_Compactness_2}
C_*= \frac{10^{( -3.19 + 0.36 \, Log \, M_d)}}{R_d}   \quad ,
\end{eqnarray}  
where, let us remind, $R_d$ is measured from photometry. 
By means of Eq. \ref{Stellar_Compactness_2}, $C_{*}$ measures, for a galaxy with a  fixed $M_d$, the deviation between the observed $R_d$  and the "expected" $R_d$ value from Eq. \ref{Stellar_Compactness_1} (obtained by using the best fit line in Fig. \ref{RD_MD}). In short, at fixed $M_d$, galaxies with the smallest $R_d$ have a high compactness ($Log \, C_* >0$), 
while galaxies with the largest $R_d$ have low compactness ($Log \, C_* < 0$).

The $Log \,C_*$ values  for  the galaxies of our sample are shown in Tab. \ref{LSB_Tab4}-\ref{LSB_Tab5} in Appendix \ref{Structural_properties} and span from -0.45 to 0.35.
\begin{figure*}
\begin{center}
\includegraphics[width=1\textwidth,angle=0,clip=true]{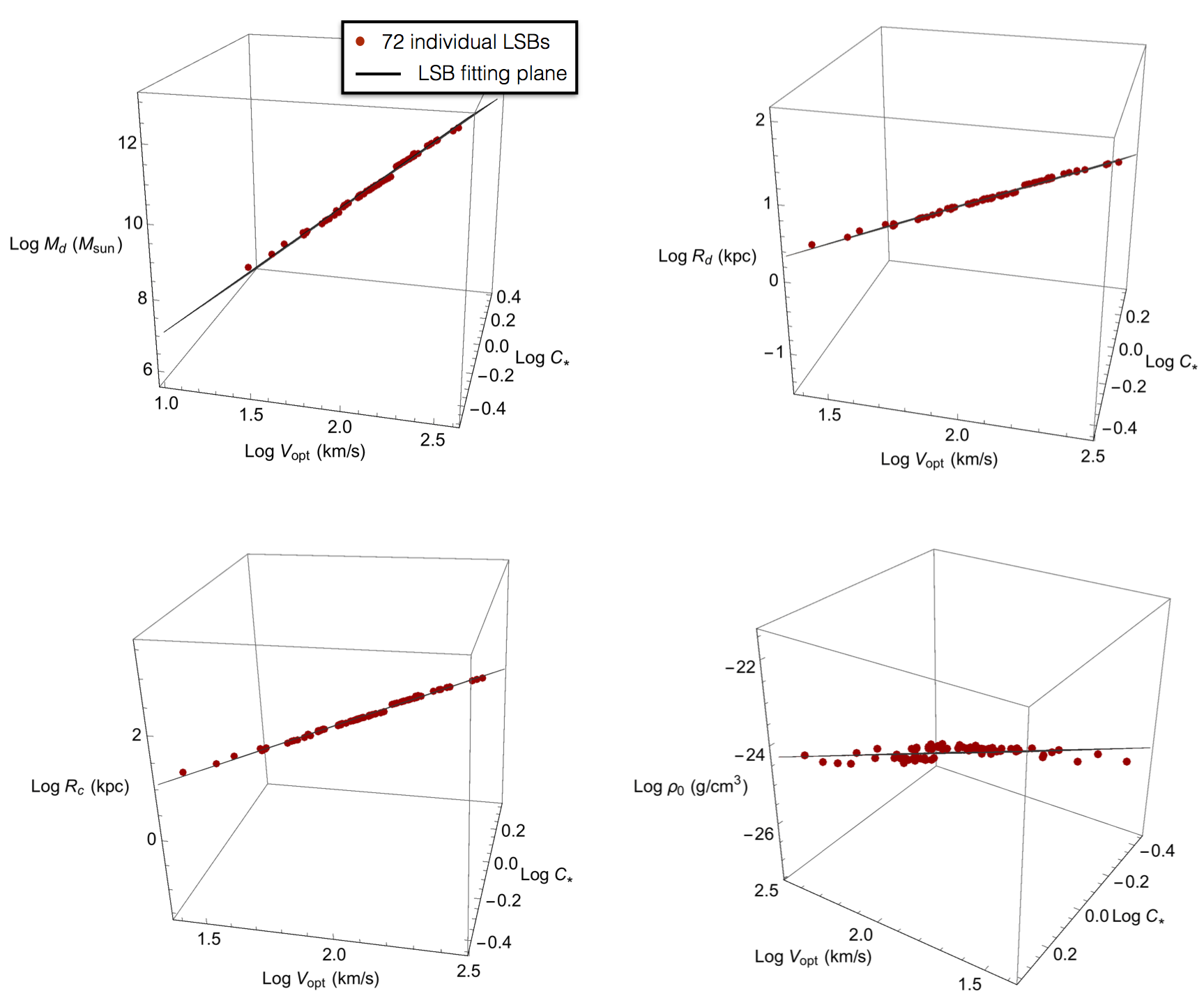}
\caption{In the panels we show  the relationships between {\it a)} the stellar disc mass, {\it b)} the stellar disc scale length, {\it c)} the DM core radius and {\it d)} the central DM core density versus the optical velocity and the compactness of the stellar distribution. The errors in the distance of galaxies, which propagates on $R_c$, $M_d$, $R_d$, $\rho_0$, are negligible in the above for 3D relationships.}
\label{Relations_Cstar}
\end{center}
\end{figure*}
\\
\\
By introducing the compactness we reduce the scatter in the relations needed to establish the analytical function of the URC-LSB in physical units. 
This is highlighted in Fig. \ref{Relations_Cstar}, where the data are shown alongside  with their best fit plane.
\begin{eqnarray}
\label{RD_Mvir_Cstar}  
Log \, M_d & = & 2.52 + 3.77 \, Log \, V_{opt} - 1.49 \, Log \, C_*  \\  \nonumber
 & &      \\  \nonumber
Log \, R_d & = & -2.27 + 1.38 \, Log \, V_{opt} - 1.55 \, Log \, C_*    \\  \nonumber
 & &      \\  \nonumber
Log \, R_c  & = & -2.62 + 1.96 \, Log \, V_{opt} - 2.20 \, Log \, C_*   \\  \nonumber
 & &      \\  \nonumber
Log \, \rho_0  & = & -20.95 - 1.84 \, Log \, V_{opt} + 3.38 \, Log \, C_*    \quad .
\end{eqnarray}  
We find that, by using Eq. \ref{RD_Mvir_Cstar}, the internal scatter of data with respect to the planes is always reduced 
 compared to the case in which $M_d$, $R_d$, $R_c$ and $\rho_0$ were expressed only 
in terms of $V_{opt}$. The previous  average scatter $\sigma \simeq 0.34$ dex of the 2D relations (Eq. \ref{MD_Vopt_fit}-\ref{Rd_Vopt_fit}), in the 3D relations (Eq. \ref{RD_Mvir_Cstar}), is reduced to $\sigma \simeq 0.06$ dex  smaller than the typical values obtained for normal spirals.  
\\
\\
We now  evaluate the analytic expression for the universal rotation curve (expressed  in physical units). 
By using Eq. \ref{V_Model} alongside with  with Eqs. \ref{V_stellar}, \ref{DM_Mass}, \ref{V_dark_matter} and 
expressing $M_d$, $R_d$, $R_c$ and $\rho_0$ as in Eq. \ref{RD_Mvir_Cstar}, 
we obtain:
\begin{eqnarray}
\label{Final_URC}  
& &V^2 (x,  \,V_{opt}, C_* )   =      2.2\, x^2  \times10^{f_1(V_{opt}, C_*)}\\  \nonumber      
                                          & &        \\  \nonumber
                                            & & \times  \, [I_0  K_0  - I_1  K_1 ]   + 1.25/x \times  10^{f_2(V_{opt}, C_*)}     \\  \nonumber
                                            & &      \\  \nonumber
                                            & & \times \{ -tg^{-1}[3.2  \, x \times10^{f_3(V_{opt}, C_*))}]      \\  \nonumber
                                            & &      \\  \nonumber
                                            & &  + \, ln[1 + 3.2 \, x  \times 10^{f_3(V_{opt}, C_*)}]       \\  \nonumber
                                            & &      \\  \nonumber
                                            & &  + 0.5 \, ln[1 + 10.24  \, x^2  \times 10^{2 \, f_3(V_{opt}, C_*)} ]  \}   \quad ,   
 \end{eqnarray}
 where $I_n, K_n$ are the modified Bessel functions evaluated at $1.6 \, x$, with $x=r/R_{opt}$ and 
 \begin{eqnarray}
\label{Final_URC_a}   
f_1(V_{opt}, C_*) & = &9.79   + 2.39 Log \, V_{opt} + 0.05 Log \, C_*       \\  \nonumber
& &      \\  \nonumber
f_2(V_{opt}, C_*) & = & -0.55   + 2.65 Log \,V_{opt} - 1.67 Log \, C_*           \\  \nonumber
& &      \\  \nonumber
f_3(V_{opt}, C_*) & = & 0.35    - 0.58 Log \, V_{opt} + 0.65 Log \, C_*    \quad .
 \end{eqnarray}
\\
\\
Finally, we plot in Fig. \ref{URCpoints} the URC (Eq. \ref{Final_URC}-\ref{Final_URC_a}) considering $Log \, C_* = 0$, corresponding to the case 
 in which all the LSBs data in Fig. \ref{RD_MD} were lying on the regression line (or, analogously, the case in which 
 the spread of LSBs data in Fig. \ref{Log_Vopt_vs_Ropt} was small). 
The curve shown in Fig. \ref{URCpoints} is in good agreement with the LSBs rotation curves data. 
On average, the uncertainty between the velocity data and the URC velocity predicted values is $\Delta V/ \,V \simeq19 \%$,  which can be reduced to 
$\Delta V/ \,V \simeq 8 \%$, when the observational errors, the systematics, the small non circularities and the prominent bulge component\footnote{The bulge component is taken into account in the coadded RCs modelling, but not in the final URC, going beyond the scope of the paper.} (as in ESO534-G020)  are taken into account in the individual RCs. This result, approximately equal to that found in normal spirals \citep{Persic_1996}, highlights the success of the URC method also in LSBs galaxies. The smallness of the  uncertainty achieved in the URC-LSB (physical) is evident is Appendix \ref{LSB rotation curves with their URC}, where the individual RCs are tested. As a gauge  we point out that  F583-4, NGC4395, UGC5005, F568V1, ESO444-G074 have a value of   $\Delta V/ \,V$  $\simeq 8 \%$.
Moreover, in Appendix \ref{LSB rotation curves with their URC}, the individual RCs are 
tested by assuming {\it i)} $Log \,C_*=0$ and {\it ii)} their values of  $C_*$ (reported in Tab.\ref{LSB_Tab4}-\ref{LSB_Tab5} in Appendix \ref{Structural_properties}).   
\begin{figure*}
\begin{center} 
\includegraphics[width=0.9\textwidth,angle=0,clip=true]{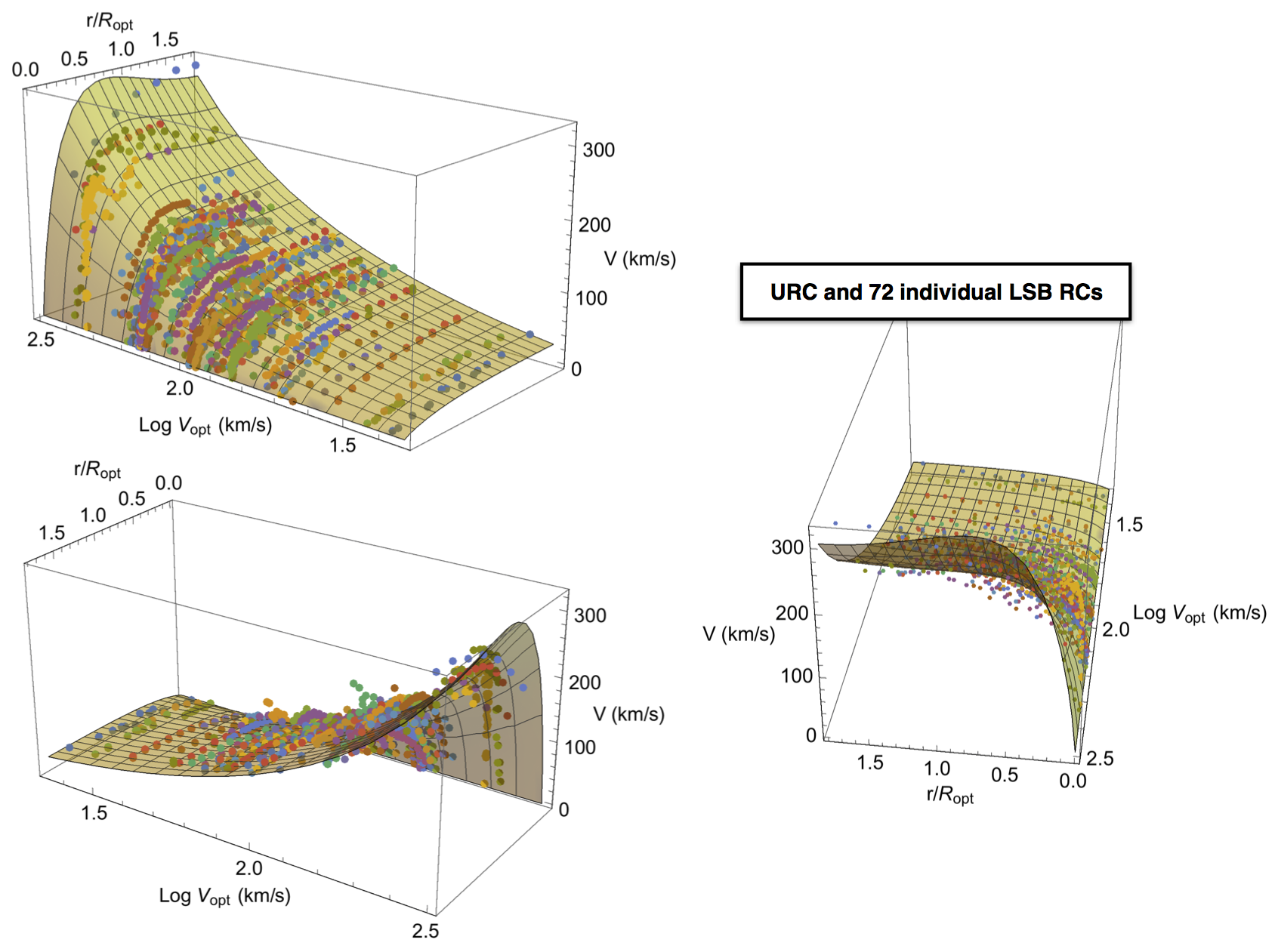}
\caption{LSBs universal rotation curve (URC), with compactness $Log \,C_*=0$, and the individual 72 LSBs rotation curves.}  
\label{URCpoints}
\end{center}
\end{figure*}

\begin{figure*}
\begin{center} 
\includegraphics[width=0.9\textwidth,angle=0,clip=true]{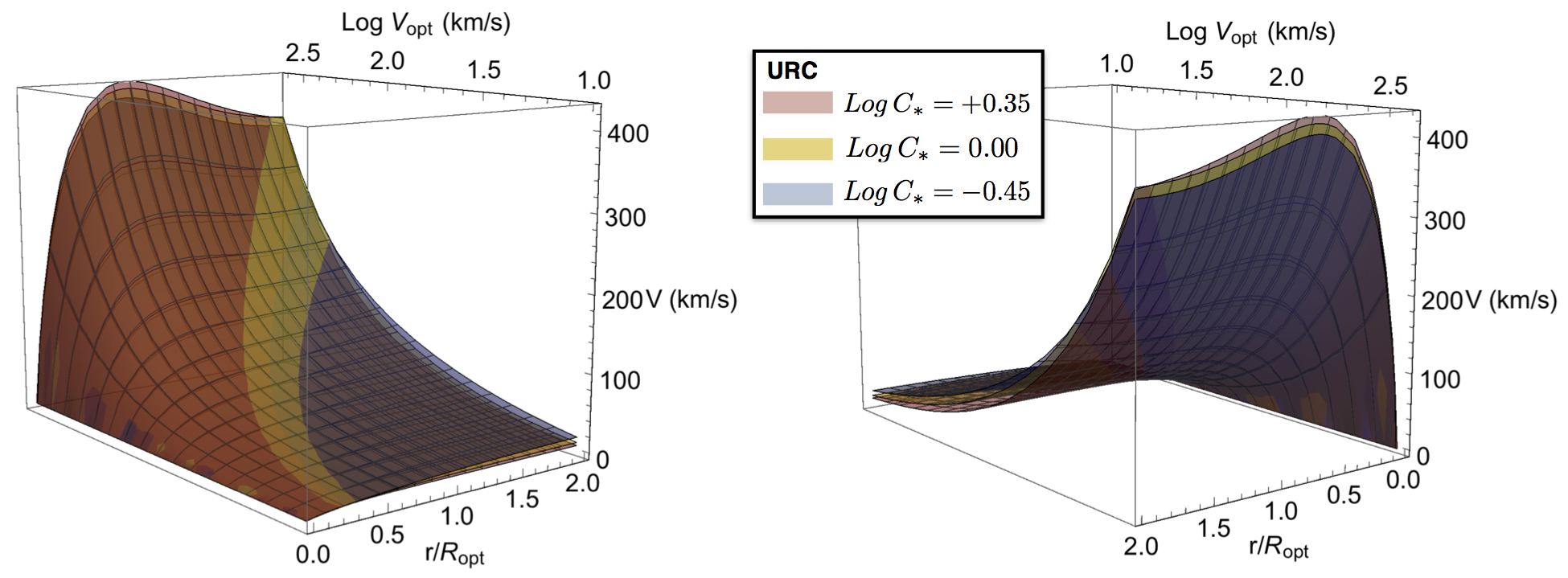}
\caption{Universal rotation curves (URC) in physical velocity units for three different values of stellar compactness; low ($Log \,C_* = -0.45$), standard 
($Log \,C_* = 0.00$) and high 
($Log \,C_* = +0.35$) stellar compactness, respectively in {\it blue, yellow} and {\it red} colors.  The figure in the second panel corresponds to that of the first panel when rotated by $180^{\circ}$ around the velocity axis.}
\label{URC_3D_Comp3}
\end{center}
\end{figure*}

\begin{figure*}
\begin{center} 
\includegraphics[width=0.9\textwidth,angle=0,clip=true]{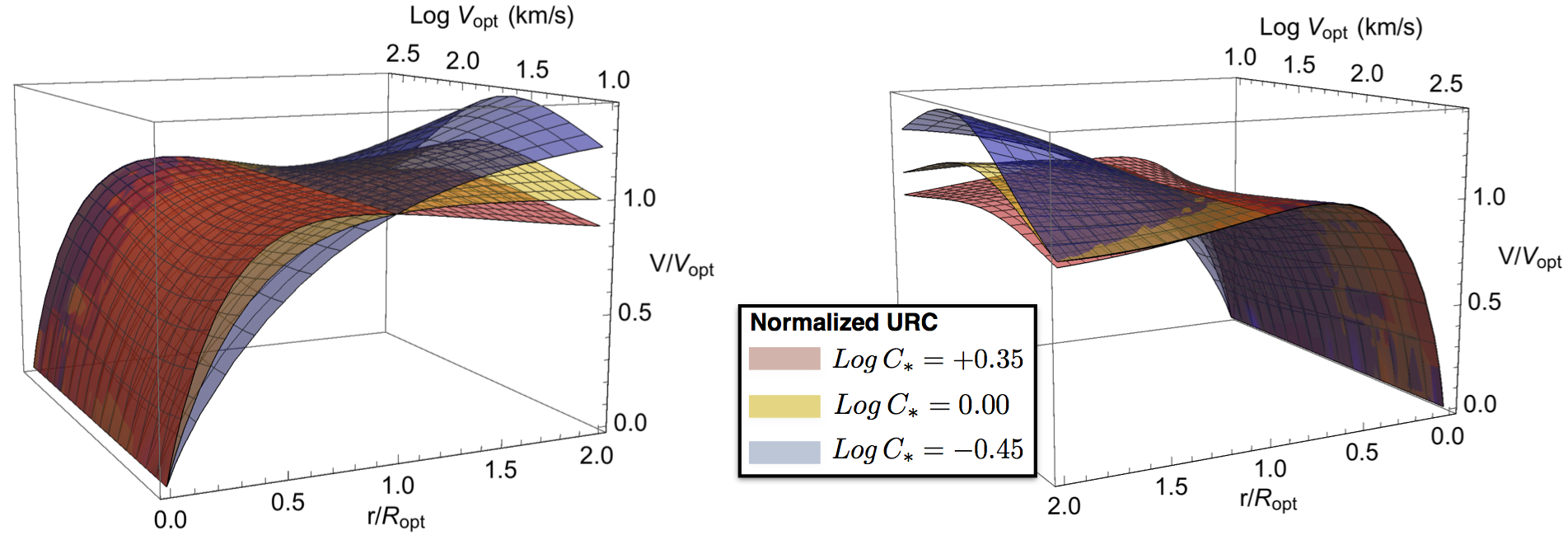}
\caption{Universal rotation curves (URC) in double normalised units for three  values of the stellar compactness: $Log \, C_* = -0.45 \, ,\, 0.00 \, , \,+0.35$, respectively in {\it blue, yellow} and {\it red} colors. Notice that the figure in the second panel corresponds to that of the first panel rotated of $180^{\circ}$ around the velocity axis.}
\label{URC_3D_Norm_Comp3}
\end{center}
\end{figure*}

Finally, in Fig. \ref{URC_3D_Comp3} we show the URC obtained with three significant different values of stellar compactness. The central yellow surface has $Log \, C_* = 0.00$ (standard case) and the other two surfaces have $Log \, C_* = -0.45$ (the minimum value achieved in the LSB sample) and $Log \, C_* = + 0.35$ (the maximum one).
The three surfaces appear similar, however when we normalise them with respect to $V_{opt}$ along the 
velocity axis, their profiles appear different. See Fig. \ref{URC_3D_Norm_Comp3}. 
Nevertheless, the differences between the  URC with $Log \, C_* = 0.00$ and the URC with the appropriate  values of  $C_*$ for each individual object lie within the URC errorbars for  most of the objects (see Appendix \ref{LSB rotation curves with their URC}).
\noindent
\noindent

\subsection{The relevance of $C_*$ in LSB galaxies}
\label{Consequences}
\noindent
Completing our analysis, we have discovered the relevance of $C_*$ in the LSB galaxies. 
By resuming, this work shows that:
\\
{\it i)} the compacteness is linked to the spread in the $V_{opt}$ - $R_d$ plot (Fig \ref{Log_Vopt_vs_Ropt}). Galaxies at fixed $V_{opt}$ can have smaller $R_d$ (higher $C_*$) 
or larger $R_d$ (lower $C_*$) than the average. The range of $Log \, R_d$ at fixed $V_{opt}$ can reach almost 1 dex; 
\\
{\it ii)} the profiles of the various RCs can be affected by the compactness (see e.g. Fig. \ref{URC_3D_Norm_Comp3}). Thus, the spread in the profiles  of the RCs in each velocity bin, (see Fig. \ref{InnerOuterSlope}-\ref{RC_Vel_bins}) is not only due to the large width of the optical velocity bins\footnote{Given the limited number of available RCs, each  optical velocity bin includes galaxies with a certain range in  $V_{opt}$, causing the corresponding  RCs to  have (moderately) different  profiles, analogously to normal spirals.}, 
but it is also due to the different values of the galaxies compactness.   
\\
{\it iii)} the compactness is a main source for the large scatter ($\sigma \simeq 0.34$) in the 2D scaling relations (see Fig. \ref{Mstar_vs_Vopt}- \ref{MD_Mvir}-\ref{Relations_Cstar}).
\\
\\
Taking all this into account, we point out that in the URC-LSB  building procedure, 
having an improved statistic,  the  optimal  approach would be  considering from the start to bin the available  RCs  in 
$C_\star$ (obtained by  the spread of data in the $V_{opt}$ - $R_d$ plot in Fig. \ref{Log_Vopt_vs_Ropt}) contemporaneously  
to  $V_{opt}$.  Moreover, with a sufficiently higher statistics, we can also increase the number of the velocity bins and characterise each of them with a smaller $V_{opt}$ range to reach the performance of \cite{Persic_1996}.    

Finally, we stress that in the LSBs there is no one-to-one correspondence among the optical velocity, the optical radius, the luminosity, the virial mass and other galaxies quantities. Then, if we order the RCs normalised in radial units, according to quantities different from the optical velocity (as  in 
Fig.\ref{RCs_3D}), they would not lie on a unique surface but,  according to  the spread of the  stellar  compactness among the objects,  will  give  rise to a spread  of RC data lying  on  different surfaces. 

\section{The correlation between the compactnesses of the stellar and the DM mass distributions}
\label{The correlation between $C_*$ and $C_{DM}$}
\noindent

Following \cite{Karukes_2017}, we evaluate also the {\it compactness of the DM halo} $C_{DM}$, 
i.e. we investigate the case in which the galaxies with the same virial (dark) mass $M_{vir}$ exhibit different core radius $R_c$. 
\begin{figure}
\begin{center}
\includegraphics[width=0.46\textwidth,angle=0,clip=true]{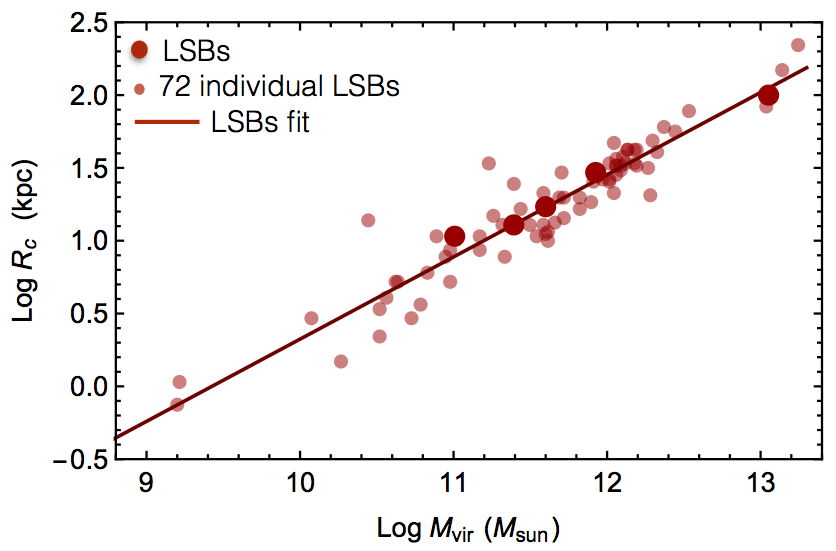} 
\caption{Relationship between the DM halo core radius and the virial mass.}
\label{C_darkmatter}
\end{center}
\end{figure}
The $M_{vir}$ vs $R_c$  relationship is shown in Fig.\ref{C_darkmatter} alongside with the best fit linear relation, 
described by 
\begin{eqnarray}
\label{DM_Compactness_1}
Log \, R_c = -5.32 + 0.56 \, Log \, M_{vir}   \quad .
\end{eqnarray}   
Then, according to \cite{Karukes_2017}, we define the compactness  $C_{DM}$ of the DM halo  as:
\begin{eqnarray}
\label{DM_Compactness_2}
C_{DM}= \frac{10^{( -5.32 + 0.56 \, Log \, M_{vir})}}{R_c}     \quad .
\end{eqnarray}   
Thus, at fixed $M_{vir}$, galaxies with smaller $R_c$ have higher compactness ($Log \, C_{DM} > 0$),
while galaxies with larger $R_c$ have lower compactness ($Log \, C_{DM} < 0$).

\begin{figure}
\begin{center}
\includegraphics[width=0.46\textwidth,angle=0,clip=true]{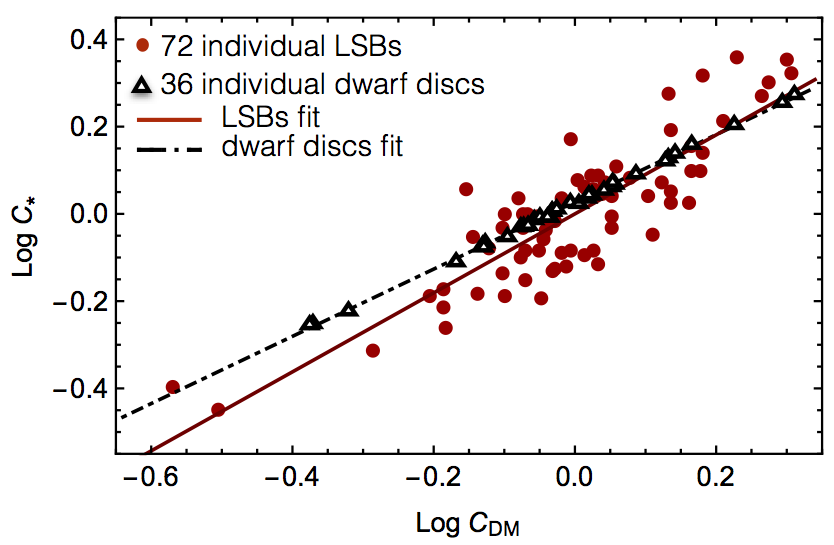} 
\caption{Relationship between the compactness of the stellar disc and the compactness of the DM halo ({\it red points}). The black triangles refer to the dwarf discs of ~\citealp{Karukes_2017}. The solid and the dot-dashed lines are the best fit relations for LSBs and dwarf discs.}
\label{Cs_Cdm}
\end{center}
\end{figure}

The values obtained for $Log \, C_{DM}$ are reported in Tab. \ref{LSB_Tab4}-\ref{LSB_Tab5} in Appendix \ref{Structural_properties} and span from -0.57 to 0.30.
\\
\\
Then, we plot the compactness of the stellar disc versus the compactness of the DM halo  in Fig. \ref{Cs_Cdm}.
We note that $C_*$ and $C_{DM}$ are strictly related: galaxies with high $C_*$, also have high $C_{DM}$. The logarithmic data are well fitted by the linear relation: 
\begin{eqnarray}
\label{DM_Compactness_3}
Log\; C_* = 0.00 + 0.90 \,Log \, C_{DM}     \quad .
\end{eqnarray}   
The results are in very good agreement with those obtained for {\it dd} galaxies ~\citep{Karukes_2017}, whose best-fitting relation is given by: $Log\; C_* = 0.77 \,Log \, C_{DM} + 0.03$. In the figure we realize that the  average difference between the two relationship is just  about 0.1 dex.

This result is remarkable because the same relation is found for two very different types of galaxies (LSBs and {\it dd}s). The strong relationship between the two  {\it compactnesses} certainly indicates that the stellar and the DM distributions follow each other very closely. In a speculative way, 
given the very different distribution of luminous matter in an exponential thin disc and the distribution of DM in a spherical cored halo, 
such strong correlation in Eq. \ref{DM_Compactness_3} might point to a non-standard interaction between the baryonic and the dark matter; a velocity dependent self-interaction in the dark sector; or a fine tuned baryonic feedback (e.g. ~\citealp{Di_Cintio_2014, Chan_2015}).

\section{Conclusions}\label{Conclusions}
\noindent
We analyzed a sample of 72 low surface brightness (LSB) galaxies selected from literature, whose optical velocities $V_{opt}$ span from $\sim 24$ km/s to $\sim 300$ km/s. Their rotation curves, normalised in the radial coordinate with respect to the stellar disc scale length $R_d$ (or the optical radius $R_{opt}$) and ordered according to the increasing optical velocity $V_{opt}$, follow a universal trend (Fig. \ref{RCs_3D}), analogously to the 
normal (high surface brightness) spirals.  
This led us to build the {\it universal rotation curve} (URC) of LSBs
as in \cite{Persic_1996},  i.e. to find an analytic expression to reproduce any  circular velocity by means of only few observable parameters (e.g. $R_{opt}$ and $V_{opt}$).    

The building of the URC allows us to obtain the properties of the stellar and DM distribution and to evaluate the scaling relations valid for the whole population of objects. 
The analysis on the  LSBs RCs leads us to a scenario which is very similar qualitatively, but not quantitatively, to that of the normal spirals.
In detail, in both cases, we observe that the main  contribution to the circular velocity, in the innermost  galactic region, is given by the stellar disc component, while, in the external region  is given by a   cored DM spherical halo. 
Moreover, the fraction of DM that contributes to the RCs is more relevant as lower $V_{opt}$ is, i.e. in smaller galaxies (Fig. \ref{Velocity_1}). 

The scaling relations among the galactic properties seem to follow a similar trend in LSB galaxies and high surface brightness (HSB) spirals 
(Fig. \ref{RC_RD_relation}-\ref{Mstar_vs_Vopt}-\ref{Rho0Sigma0}).  

On the other side, there is a clear difference: we realise the presence of a large scatter in the LSBs relationships with respect to that found in normal spirals (see \citealp{Lapi_2018}).
Such difference can be traced back to the large spread of the $V_{opt}$-$R_d$ data (see Fig. \ref{Log_Vopt_vs_Ropt}) or, analogously, to 
the large spread of the $R_d$-$M_d$ relationship in Fig. \ref{RD_MD}.  This  finding leads us to introduce the concept of {\it compactness} of the  luminous matter distribution $C_*$, involved for the first time in 
\cite{Karukes_2017} to cope with a similar issue in the case of dwarf disc ({\it dd}) galaxies. 

We have that in galaxies with a fixed value for $M_d$, the smaller 
is $R_d$, the higher is $C_*$.
By considering $C_*$ in the scaling relations, the scatter is much reduced (it becomes smaller than that of the normal spirals). By involving this new parameter, we proceed with the building of the analytic universal expression to describe 
all the LSBs rotation curves (in physical units, km/s vs kpc). 
The resulting URC, $V(r; R_{opt}, \, V_{opt}, \, C_*)$ in Eq. \ref{Final_URC}-\ref{Final_URC_a}, well describes all the rotation curves of our sample (Fig. 
\ref{URCpoints}-\ref{Single_URC}-\ref{Single_URC_1}-\ref{Single_URC_2}-\ref{Single_URC_3}-\ref{Single_URC_4}). The average scatter of the RCs data from the fitting surface in Fig. \ref{URCpoints} achieves the small value of  $\Delta V /\, V \simeq 0.08$, taking into account the observational errors, the systematics and the small non-circularities in the motion. This result remarks the success of the method leading to the URC and of the relevance of $C_*$ in the RCs profiles (Fig. \ref{URC_3D_Norm_Comp3}) and in the scaling relations, which has been discovered in building the URC. 

With larger statistics, one should subdivide the RCs according to the galaxies $C_*$ and $V_{opt}$.

An important finding concerns the compactness of the DM distribution $C_{DM}$, indicating galaxies with the 
same virial mass and different core radius (Fig. \ref{C_darkmatter}).
We find a strong correlation between $C_*$ and $C_{DM}$ as also found in \cite{Karukes_2017} (Fig \ref{Cs_Cdm}): 
the {\it distributions of stellar disc and of its enveloping   dark matter halo are entangled}. In a speculative way, this finding appears to be of very important relevance for 
the nature of DM. In fact, 
the strong correlation between $C_*$ and $C_{DM}$ may hint to the existence of non standard interactions between the luminous matter  and the 
dark matter, or non trivial self-interaction in the DM sector or a (hugely) fine-tuned baryonic feedback on the collisionless DM distribution.

Finally, the LSBs URC provides us with  the best  observational data to
test specific density profiles (e.g. NFW, WDM, Fuzzy DM) or
alternatives to dark matter (e.g.  MOND). The normal spirals' URC, in
connection with the normal spirals' $R_{opt} $ vs $V_{opt}$ relationship,
is a function of $V_{opt}$: $V_{URC \,(ns)}(r/R_{opt},V_{opt})$ ({\it ns} stands for normal spirals).
Therefore,  to  represent {\it all} the normal spirals' individual  RCs it is sufficient
to evaluate $V_{URC \,(n s)}(r/R_{opt},V_{opt})$ for a reasonable
number $j$ of $V_{opt}$ values, homogeneously  spread in
the spirals  $V_{opt}$ range. Any mass model under test must
reproduce the $V_{opt}$ dependent URC. Instead, the LSBs URC,
in connection with the LSBs $R_{opt} $ vs $V_{opt}$ and $C_*$
relationship, is a function of  {\it two} galaxy structural properties: $V_{opt}$ and $C_*$. In this
case, to represent {\it all} the LSBs RCs we have  to build
$V_{URC \,(LSB)}(r/R_{opt},V_{opt}, C_*)$. We need  
a large sample of RCs of galaxies of different $V_{opt}$ and $C_*$ yielding  
a reasonable number of  RCs in  each of the more numerous ($V_{opt};\, C_*$) bins we have to employ.   
The galaxies model under test must reproduce a much complex (observational driven) URC  than that of normal  spirals 
which depends on just the structural parameter $V_{opt}$.

\section*{acknowledgments}

\noindent 
We thank E. Karukes and A. Lapi for useful discussions. We thank Brigitte Greinoecker for improving the text. We thank the referee for her/his several inputs that have improved the paper.
%\end{acknowledgments}
%
%

%\bibliographystyle{unsrt}
%\bibliography{Bibliografia}
%
\bibliography{Bibliografia}
\newpage

\begin{appendix}
%\label{appendix}
%
%
%
\section{Universality in normal spirals}
\label{RC_Normal_Spirals}
\begin{figure*}
\begin{center} 
\includegraphics[width=0.4\textwidth,angle=0,clip=true]{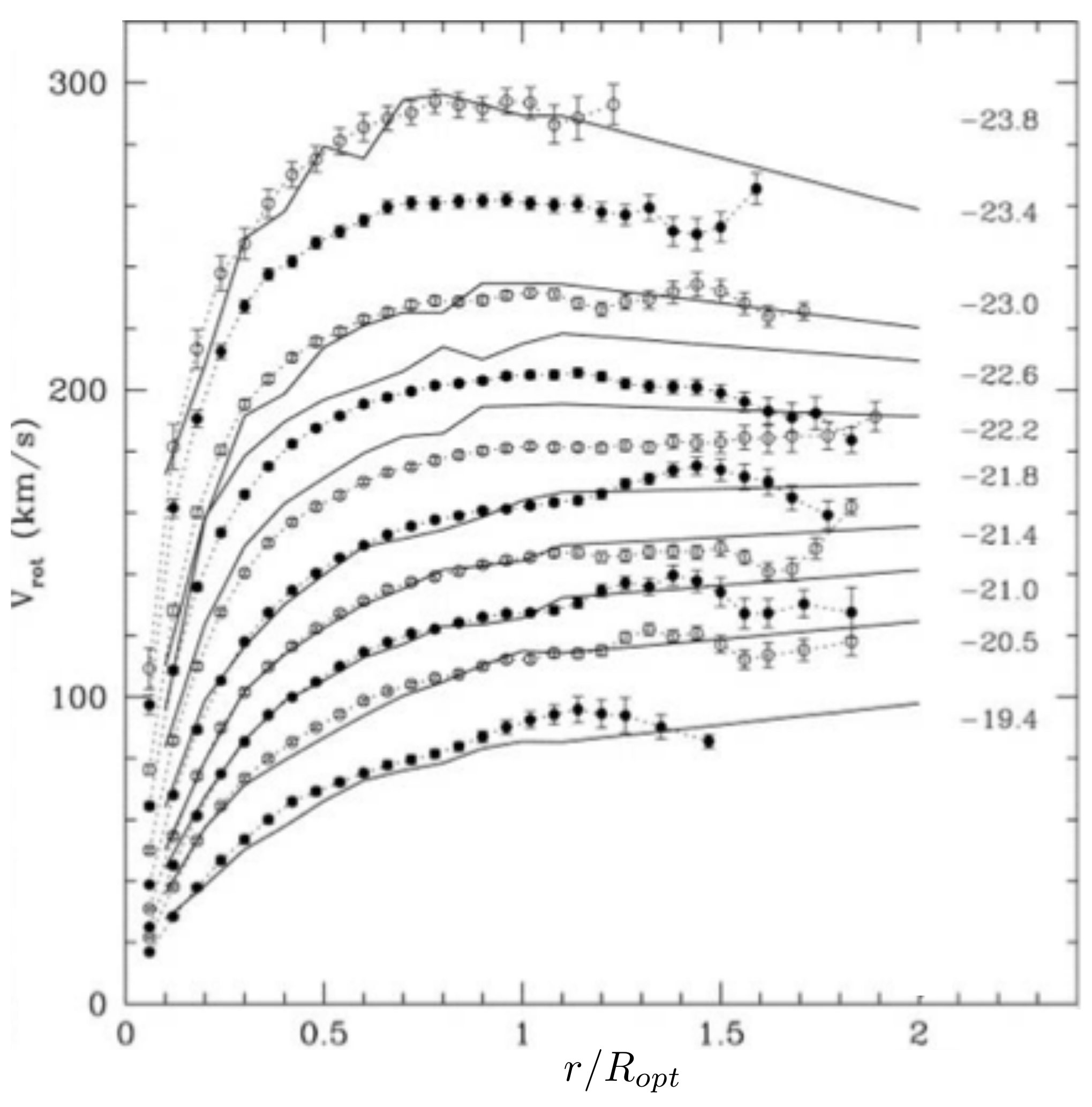}
\includegraphics[width=0.50\textwidth,angle=0,clip=true]{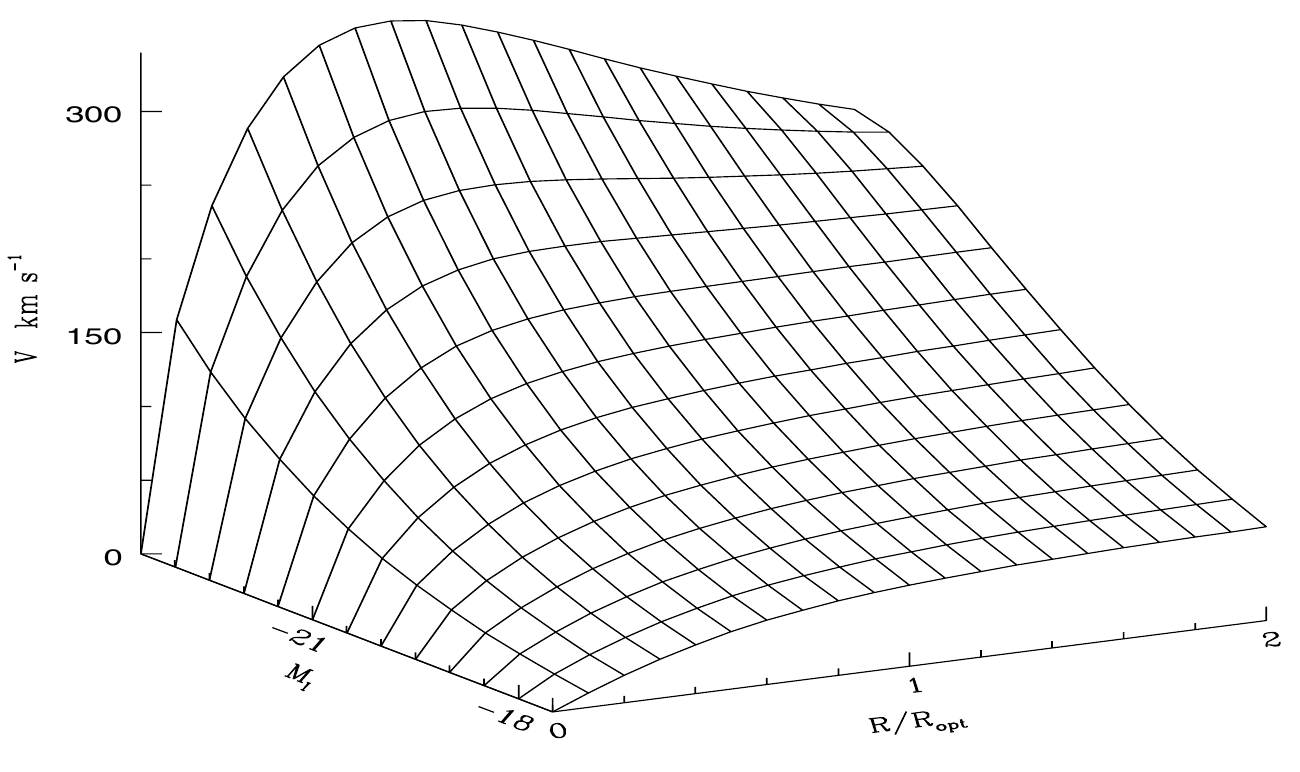}
\caption{{\it Top panel}: coadded rotation curves from 3100 normal spirals, obtained by plotting together the results by \citealp{Persic_1996} and \citealp{Catinella_2006} (originally in the slides by \citealp{slides}). Also indicated the absolute I-magnitudes.  {\it Bottom panel}: universal rotation curve (URC) \citep{Persic_1996}.}  
\label{RC_Salucci}
\end{center}
\end{figure*}

\noindent
 Fig. \ref{RC_Salucci}, from \cite{Persic_1996} and \cite{Catinella_2006}, allows us to appreciate the {\it universality} of the rotation curves in normal spirals after the radial normalisation. Let us point out the trend of the RCs from small to large galaxies.
\section{LSB galaxies sample and references}
\label{LSBs sample and references}
\noindent
In Tab. \ref{LSB_References_a}, we report the list of the LSB galaxies of this work with their related references.   
\begin{table*}
\begin{tabular}{p{2.5cm}p{6.0cm}p{2.5cm}p{6.0cm}}
\hline
Galaxy      &  Reference     &       Galaxy      &  Reference     \\
\hline
NGC 100  &       \cite{deBlok_2002}     &  UGC 11557  &    \cite{Swaters_2003}   \\
NGC 247	  &      \cite{Carignan_1990}  &    UGC 11583  &    \cite{deBlok_2001}  \\
NGC 959  &	  \cite{Kuzio_de_Naray_2008}   &  UGC 11616  &    \cite{deBlok_2001}   \\  
NGC 2552  &	  \cite{Kuzio_de_Naray_2008}   &   UGC 11648  &    \cite{deBlok_2001}  \\
NGC 2552   &	    \cite{deBlok_2002}  &           UGC 11748  &   \cite{deBlok_2001}   \\
NGC 2552  &	  \cite{Swaters_2003}  &     UGC 11819  &    \cite{deBlok_2001}  \\
NGC 2552  &	  \cite{van_den_Bosch_2001} &  ESO 186-G055  &      \cite{Pizzella_2008}  \\
NGC 3274   &	    \cite{deBlok_2002} &   ESO 206-G014  &     \cite{Pizzella_2008}  \\
NGC 3274  &	 \cite{Swaters_2003}  &   ESO 215-G039  &      \cite{Palunas_2000}  \\
NGC 3347B  &	  \cite{Palunas_2000} &    ESO 234-G013  &     \cite{Pizzella_2008}    \\
NGC 4395  &        \cite{deBlok_2002} &     ESO 268-G044  &      \cite{Palunas_2000}  \\
NGC 4395  &    \cite{van_den_Bosch_2001} &  ESO 322-G019  &     \cite{Palunas_2000}  \\
NGC 4455  &     \cite{deBlok_2002} &     ESO 323-G042  &      \cite{Palunas_2000}   \\
NGC 4455  &    \cite{Marchesini_2002}   &   ESO 323-G073  &      \cite{Palunas_2000}    \\
NGC 4455  &    \cite{van_den_Bosch_2001} &    ESO 374-G003  &      \cite{Palunas_2000}    \\
NGC 5023  &    \cite{deBlok_2002} &         ESO 382-G006  &     \cite{Palunas_2000}  \\
NGC 5204  &    \cite{Swaters_2003}  &       ESO 400-G037  &      \cite{Pizzella_2008}  \\
NGC 5204  &    \cite{van_den_Bosch_2001}  &     ESO 444-G021  &      \cite{Palunas_2000}  \\
NGC 7589  &    \cite{Pickering_1997}   &     ESO 444-G047  &      \cite{Palunas_2000}   \\
UGC 628    &     \cite{deBlok_2002}  &          ESO 488-G049  &      \cite{Pizzella_2008}    \\
UGC 634  &       \cite{vanZee_1997} &           ESO 509-G091  &     \cite{Palunas_2000}  \\
UGC 731  &         \cite{deBlok_2002}  &         ESO 534-G020  &      \cite{Pizzella_2008}  \\
UGC 731  &      \cite{Swaters_2003}    &         F561-1  &    \cite{deBlok_1996}   \\
UGC 731  &     \cite{van_den_Bosch_2001} &    F563-V1  &    \cite{deBlok_1996}  \\
UGC 1230  &     \cite{deBlok_2002}   &             F563-V2  &   \cite{KuziodeNaray_2006}   \\
UGC 1230  &    \cite{vanderHulst_1993}   &    F563-V2  &    \cite{deBlok_1996}   \\
UGC 1281  &   \cite{KuziodeNaray_2006}    &     F565-V2  &    \cite{deBlok_1996}  \\
UGC 1281  &      \cite{deBlok_2002}  &                 F568-1  &    \cite{Swaters_2000}    \\
UGC 1551  &    \cite{Kuzio_de_Naray_2008}    &   F568-3  &   \cite{KuziodeNaray_2006}  \\
UGC 2684  &    \cite{vanZee_1997}    &                 F568-3  &    \cite{deBlok_2001}   \\
UGC 2936  &   \cite{Pickering_1999}   &              F568-3  &    \cite{Swaters_2000}    \\
UGC 3137  &     \cite{deBlok_2002} &                  F568-6  &    \cite{Pickering_1997}   \\
UGC 3174  &    \cite{vanZee_1997}  &                 F568-V1  &    \cite{Swaters_2000}   \\
UGC 3371  &     \cite{deBlok_2002} &                  F571-8  &    \cite{Marchesini_2002}    \\
UGC 3371  &    \cite{van_den_Bosch_2001} &     F571-8  &   \cite{deBlok_2001}   \\
UGC 4115  &        \cite{deBlok_2001}  &               F571-V1  &    \cite{deBlok_1996}     \\
UGC 4278  &    \cite{deBlok_2002}  &                   F574-1  &    \cite{Swaters_2000}     \\
UGC 5005  &    \cite{deBlok_a_1997}  &               F574-2  &    \cite{deBlok_1996}    \\
UGC 5272  &    \cite{Kuzio_de_Naray_2008}    &    F579-V1  &    \cite{deBlok_2001}    \\
UGC 5272  &    \cite{deBlok_2002}   &             F583-1  &    \cite{Kuzio_de_Naray_2008}    \\
UGC 5716  &    \cite{vanZee_1997} &            F583-1  &    \cite{Marchesini_2002}    \\
UGC 5750  &    \cite{KuziodeNaray_2006}  &      F583-1  &    \cite{deBlok_2001}   \\
UGC 5750  &    \cite{deBlok_2002}   &             F583-1  &    \cite{deBlok_1996}     \\
UGC 5999  &    \cite{vanderHulst_1993}  &      F583-4  &    \cite{KuziodeNaray_2006}  \\
UGC 7178  &   \cite{vanZee_1997}  &              F583-4    &  \cite{deBlok_2001}    \\
UGC 8837  &    \cite{deBlok_2002}   &             F730-V1  &    \cite{deBlok_2001}   \\
UGC 9211    &    \cite{van_den_Bosch_2001}   &    PGC 37759  &    \cite{Morelli_2012}    \\
UGC 11454  &    \cite{deBlok_2001}   &              &             \\
\hline 
\end{tabular}
\caption{LSB sample: galaxy names and references of their RCs and photometric data. 
Note that some galaxies have multiple rotation curve data.}
\label{LSB_References_a}
\end{table*}
\section{Rotation curves in physical units}
\label{Rotation_curves_in_physical_units}
\begin{figure}
\begin{center} 
\includegraphics[width=0.45\textwidth,angle=0,clip=true]{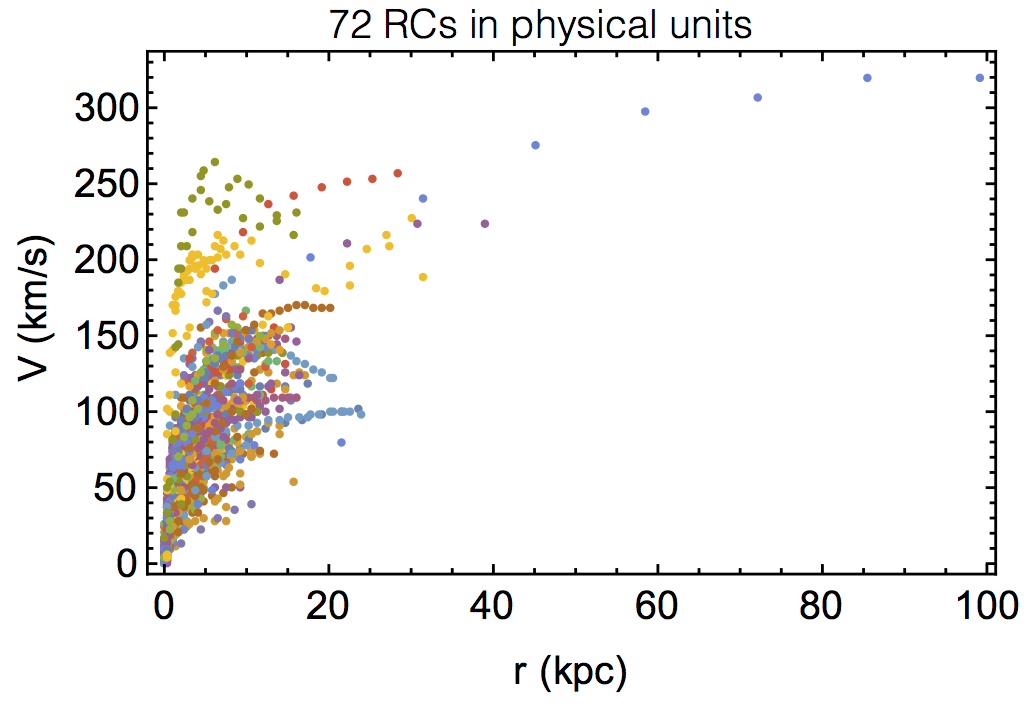} 
\caption{The 72 LSB RCs in physical units (all data).}  
\label{Vr_complete}
\end{center}
\end{figure}

\noindent
In Fig. \ref{Vr_complete}, the 72 LSB RCs are shown in physical units. Here, all the data are included, while, the first panel of Fig. \ref{RC_together} includes only data with $r\leq 30$ kpc.

\section{Rotation curves in velocity bins}
\label{Double_norm_RC_in_Vbin}
\begin{figure*}
\begin{center} 
\includegraphics[width=0.6\textwidth,angle=0,clip=true]{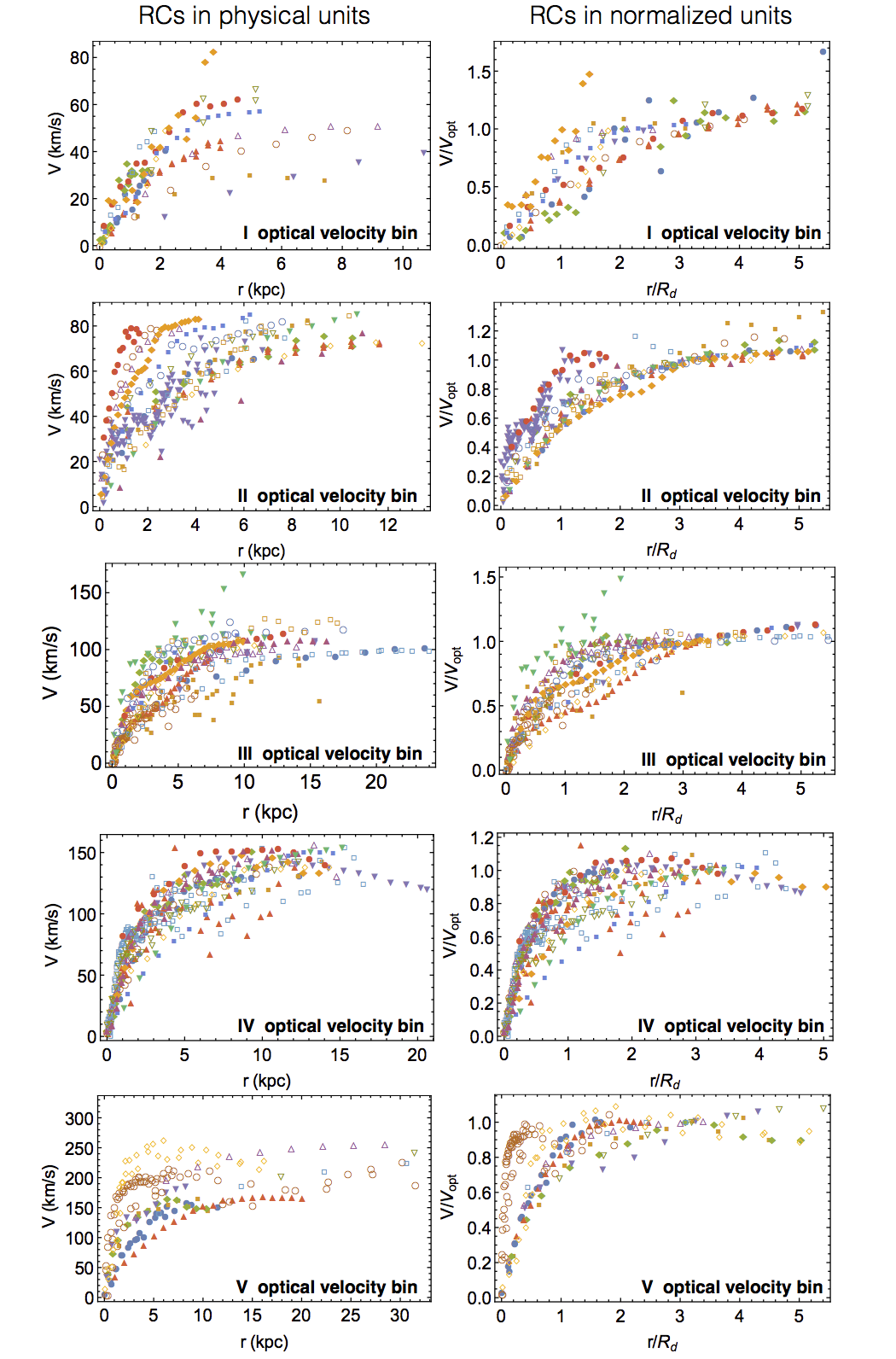}
\caption{LSBs rotation curves belonging to each of the five optical velocity bins.}  
\label{RC_Vel_bins_b}
\end{center}
\end{figure*}

\noindent
In Fig. \ref{RC_Vel_bins_b} we show  the LSBs rotation curves  separately in the five velocity bins, both in physical units and in double normalised units (i.e. along the radial and the velocity axes). 
%
%
%
%
%
%\vfill\eject
\section{Construction of the coadded rotation curves}
\label{Radial_binning}
\noindent
In Tab. \ref{LSB_Tab2a}-\ref{LSB_Tab2b} we show the data related to the five coadded rotation curves obtained in Section \ref{The coadded RC of LSB galaxies}. In detail, the first column describes the center of the radial bins represented
by the coloured points in Fig. \ref{RC_Vel_bins}-\ref{Binning_Double_Norm}-\ref{Velocity_1}. The second column indicates the number of RCs data
(grey points in Fig. \ref{RC_Vel_bins}) belonging to each radial bin. Finally, the third and the fourth columns show the velocity data and their error-bars in physical units,
related to the coadded RCs in the second panel of Fig. \ref{Binning_Double_Norm} 
and in Fig. \ref{Velocity_1}.
\begin{table}
\begin{tabular}{p{2.2cm}p{1.5cm}p{1.5cm}p{1.5cm}}
\hline
$r/R_D$   &    N. data  &       V     &     ErrorBar \\
                &                  &   km/s    &     km/s    \\
  (1)          &       (2)      &       (3)     &      (4)         \\     
\hline            
\hline
I  velocity bin   &            &                &                      \\
\hline
0.2           &        15       &    7.3     &    1.2        \\
0.6           &        19       &     16.7  &    1.7         \\
 1.0          &       18        &      25.6  &    2.4       \\
 1.4          &        19       &     34.2   &   3.1        \\
1.8           &        13       &     37.1    &   1.6       \\
2.25         &        13       &    42.4     &  1.4        \\
2.75         &        10       &     41.9    &   2.1       \\
3.25         &        13       &    45.5    &   0.9        \\
3.75         &         5        &   48.4     &  0.7         \\
4.25         &         5        &    51.1    &  1.4         \\
4.75         &         5        &   49.9     & 1.1          \\
5.25         &         5        &  56.4      &  4.2          \\
\hline
II  velocity bin   &            &                &                      \\
\hline
0.2           &      62        &     25.0    &  1.5      \\
 0.6           &      70        &    40.0     &  1.3     \\
 1.0           &      39         &     52.0    &  1.9      \\
 1.4           &       26        &    56.5     &  1.9       \\
1.8            &       26        &    62.3      &  1.2     \\
2.25          &       23        &    64.8      &  1.3      \\
2.75          &       23         &    70.7     &  0.8       \\
 3.25         &       16         &     74.3    &   0.5      \\
 3.75         &       15         &  76.3       &   1.2      \\
4.25          &       12         &  78.6       &  1.4      \\
4.75          &       12         &    81.0     &  1.6       \\
5.25          &        9          &    81.7      &   2.1    \\
\hline
III  velocity bin    &            &                &                      \\
\hline
0.2            &     86          &     25.3      &   1.8      \\
0.6            &      56         &     53.7      &   1.9      \\
1.0            &       46        &     71.8      &   2.7     \\
1.4            &      45         &     81.1      &   2.8     \\
1.8            &      35         &     89.9      &  3.2       \\
2.25          &       39        &     93.6      &  1.3       \\
2.75          &       29         &    97.2       &  1.7      \\
3.25          &       20         &  101.3      &  0.5       \\
3.75          &       10         &   104.0     &   0.8       \\
4.25          &      8            &    106.9    & 1.0         \\
4.75          &      10          &  107.8      &   1.4        \\
5.25          &       6           &    107.9    &  2.0        \\
\hline
\end{tabular}
\caption{Rotation curves for each optical velocity bin of LSB galaxies. Columns: (1) center of the radial bin; (2) number of data in each bin; (3) coadded velocity for each bin; (4) velocity error. 
In order to express the radial coordinate in physical units, the data of the first column relative to each velocity bin must be multiplied by the respective average value of disc scale length 
$\langle R_D \rangle$, reported in Tab. \ref{LSB_Tab2}.}
\label{LSB_Tab2a}
\end{table}
\begin{table}
\begin{tabular}{p{2.2cm}p{1.5cm}p{1.5cm}p{1.5cm}}
\hline
$r/R_D$   &    N. data  &       V     &     ErrorBar \\
                 &                 &   km/s     &     km/s    \\
  (1)          &       (2)      &       (3)     &      (4)         \\     
\hline            
\hline
IV  velocity bin     &            &                &                      \\
\hline      
 0.2          &     141          &     47.9          &     2.2         \\
 0.6          &     81          &      90.4          &     2.0         \\
 1.0           &     54          &     112.2          &     2.6         \\
 1.5          &     58          &      121.8          &     2.2         \\
 2.1          &     41          &     128.6          &     3.1         \\
 2.7          &     28          &      133.7          &     2.9         \\
 3.3          &     17          &      136.0          &     2.5         \\
 3.9          &     9          &     138.9          &     3.0         \\
 4.7          &     8          &     129.5          &     2.8           \\
\hline
V  velocity bin     &            &                &                      \\
\hline
  0.2         &      71         &       127.1         &      7.2        \\
 0.6         &      32         &       148.7         &      6.1        \\
 1.         &      23         &       173.9         &      3.5        \\
 1.4         &      14         &       197.6         &      3.7        \\
 1.8         &      16         &       194.8         &      4.9        \\
 2.25         &      14         &      198.2         &      3.4        \\
 2.75         &      5         &      199.3         &      5.2        \\
 3.25         &      9         &       205.5         &      1.5        \\
 3.75         &      6         &       203.2         &      4.0        \\
 4.4         &      8         &      199.6         &      5.3        \\
 5.2         &      5         &      195.2         &      6.9       \\
\hline
\end{tabular}
\caption{It continues from Tab. \ref{LSB_Tab2a}.}
\label{LSB_Tab2b}
\end{table}
%
%

%\newpage
\section{The gas component in the Rotation Curves}
\label{Gas_component}
\begin{figure*}
\begin{center} 
\includegraphics[width=0.8\textwidth,angle=0,clip=true]{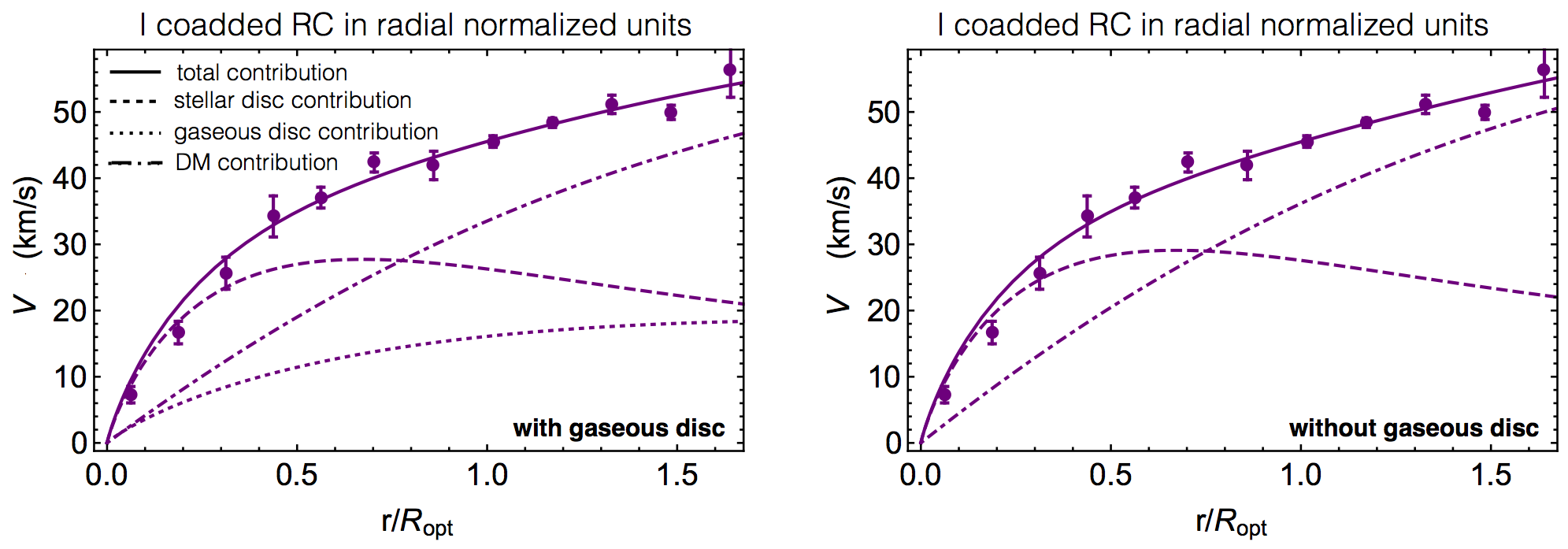}
\end{center} 
\caption{I velocity bin rotation curve best-fitted with gas ({\it left panel}) and without gas ({\it right panel}).  The {\it dashed, dot-dashed, dotted} and {\it solid lines} stand for the stellar disc, the DM halo, the gaseous disc and the total contributions to the rotation curve.}
\label{I_velocity_bin}
\end{figure*}

\noindent
The gas disc component in galaxies is an additional  component to the stellar disc and the DM halo giving a contribution to the circular velocities. At any rate, by performing a suitable test, it is possible to realize  that
the gas is (moderately) important only in the first optical velocity bin, where, in any case,  in the inner regions  the stellar component overcomes the gaseous one, while in the external region the DM component overcomes the gaseous one; thus,  the gas component  gives a modest contribution to the RC.
In Fig. \ref{I_velocity_bin}, for the first velocity bin co-added rotation curve, we compare  the mass-velocity model fit that includes the contribution from a  HI disc with the  velocity-mass model which does not. The estimated masses  of the stellar disk and of the DM halo show, in the two cases,  only a moderate change .

By modelling  the co-added rotation curve of the first $V_{opt}$ bin by means of  the stellar/HI disc +  DM halo model we get:

$M_d = 8.0 \times10^8 M_{\odot}$  ;  

$r_0 = 10.7 \, kpc$  ;      

$\rho_0=  3.2 \times 10 ^{-3} M_{\odot}/pc^3$   ;
        
$M_{vir} = 8.2\times10^{10} M_{\odot}$ ;   

$M_{HI} = 1.0 \times 10^9  M_{\odot}$.           
\\
\\
By removing the gaseous disc, we get:

$M_d = 8.8 \times10^8 M_{\odot}$       ;         

        $ r_0 = 10.7 \, kpc$ ;   
        
         $\rho_0= 3.7 \times 10 ^{-3} M_{\odot}/pc^3$   ;    
         
           $  M_{vir} = 1.0 \times 10^{11} M_{\odot}$  .
\\
\\
We remind that  $M_d, \, r_0, \, \rho_0 ,\, M_{HI}$ (all quantities inferred by the fit) are the stellar disc mass, the DM halo core radius, 
the central core mass density, the HI gaseous disc mass (including the correction for the helium contribution), respectively.  $M_{vir}$ is the virial mass. The differences in the values of  $M_d, \, r_0, \, \rho_0 , M_{vir}$, when we include gas or we  exclude the gaseous component, are inside the error bars reported in Tab. 
\ref{LSB_Tab3} related to the fit without the HI disk.

\section{Structural properties of LSB galaxies}
\label{Structural_properties}
\noindent
In Tab. \ref{LSB_Tab4}-\ref{LSB_Tab5} we report: the names of the LSB galaxies in our sample alongside their distances $D$, the stellar disc scale lengths $R_d$ and the  optical velocities $V_{opt}$ (all taken from literature). Furthermore, the table shows the values of the stellar disc mass $M_d$, the DM core radius $R_c$, the central density of the DM halo $\rho_0$, the virial mass $M_{vir}$, the central surface density $\Sigma_0 = \rho_0 \, R_c$, the compactness of the stellar mass distribution $C_*$ and that  of the DM mass distribution $C_{DM}$, all evaluated in this work.
\begin{table*}
\begin{tabular}{p{2.0cm}p{1.1cm}p{1.1cm}p{1.1cm}p{1.1cm}p{0.9cm}p{1.2cm}p{1.1cm}p{1.1cm}p{1.1cm}p{1.1cm}}
\hline
Name         &  $D$    &$ R_d  $      &     $V_{opt}$        &         $M_d $              &    $ R_c $      &    $Log \, \rho _0$     &    $M_{vir}$          &       $Log \Sigma _0$           &       $Log C_* $        &       $Log C_{DM}$   \\
     \;            &    Mpc     &     kpc            &          km/s           &  $10^7 \,M_{\odot}$   &         kpc        &    $g/ cm^3$     &    $10^9 \,M_{\odot}$    &      $M_{\odot} / pc^2$        &                 \;                 &               \;                 \\
   (1)           &         (2)             &            (3)             &          (4)                     &         (5)          &           (6)                 &            (7)               &                  (8)                      &             (9)                 &              (10)         &    (11)     \\ 
\hline
 UGC4115     & 7.8 &   0.4  &   24.2  &   6.3  &    1.1  &     -23.57    &     1.6    &     1.63    &   0.06  &     -0.15  \\
 F563V1        &    51.0    &   2.4  &   27.3  &   48  &    14  &     -25.30    &     27    &     1.01    &   -0.40  &     -0.57  \\
 UGC11583   &    5.9        & 0.3  &   27.9  &   6.5  &   0.7  &     -23.17    &     1.6    &     1.88    &   0.17  &     -0.00  \\
 UGC2684    &     8.2         &  0.8  &   36.7  &   29  &   2.9  &     -23.95    &     12    &     1.69    &   -0.00  &   -0.10  \\ 
 F574-2        &      66.0    & 4.5  &   40.0  &   192  &    33  &     -25.57    &     171    &     1.13    &   -0.45  &     -0.50  \\
 F565V2       &    36.0      &  2.0  &   45.2  &   110  &   11  &     -24.69    &     76    &     1.51    &   -0.19  &     -0.21  \\
 UGC5272    &    6.1      & 1.2  &   48.8  &   77  &   5.2  &     -24.11    &     42    &     1.77    &   -0.02  &     -0.04  \\
 UGC8837    &    5.1   &1.2  &   49.6  &   79  &   5.2  &     -24.10    &     44    &     1.78    &   -0.02  &     -0.03  \\
 F561-1        &   63.0      & 3.6  &   50.8  &   250  &    25  &     -25.15    &     244    &     1.41    &   -0.31  &     -0.28  \\
 UGC3174   &    11.8    & 1.0  &   51.7  &   72  &   4.0  &     -23.88    &     36    &     1.89    &   0.04  &     0.04  \\
 NGC4455   &    6.8        & 0.9  &   53.0  &   68  &    3.4  &     -23.75    &     33    &     1.96    &   0.08  &     0.08  \\
 UGC1281   &     5.5    & 1.7  &   55.0  &   138  &   8.5  &     -24.36    &     96    &     1.74    &   -0.08  &     -0.05  \\
 UGC1551   &    20.2         & 2.5  &   55.8  &   211  &   15  &     -24.73    &     182    &     1.61    &   -0.18  &     -0.14  \\
 UGC9211   &   12.6   &  1.3  &   61.9  &   165  &   5.9  &     -24.10    &     66    &     1.84    &   0.06  &     0.01  \\
 F583-1       &  1.6   &1.6  &   62.0  &   201  &    7.8  &     -24.29    &     90    &     1.77    &    0.00  &     -0.03  \\
 UGC5716   &     24.1    & 2.0  &   66.4  &   288  &    11  &     -24.45    &     150    &     1.75    &   -0.03  &     -0.04  \\
 UGC7178   &    24.0    &  2.3  &   69.9  &   367  &   13  &     -24.54    &     210    &     1.74    &   -0.06  &     -0.04  \\
 ESO400-G037  &   37.5   & 4.1  &   69.9  &   651  &    29  &     -25.09    &     502    &     1.55    &   -0.21  &     -0.18  \\
 NGC3274    & 0.47  & 0.5  &   68.0  &   75  &   1.5  &     -23.01    &     18    &     2.33    &   0.35  &     0.30  \\
 F583-4         &   49.0   &  2.7  &   70.5  &   438  &   16  &     -24.69    &     275    &     1.69    &   -0.10  &     -0.08  \\
 F571V1        &  79.0   & 3.2  &   72.4  &   549  &    21  &     -24.83    &     382    &     1.66    &   -0.14  &     -0.10  \\
 NGC5204     &    4.9   & 0.7  &   73.1  &   115  &    2.2  &     -23.24    &     33    &     2.27    &   0.30  &     0.27  \\
 UGC731       &   8.0    & 1.7  &   73.3  &   298  &   8.5  &     -24.20    &     147    &     1.90    &   0.04  &     0.05  \\
 NGC959      &   7.8     &  0.9  &   75.3  &   172  &   3.6  &     -23.57    &     60    &     2.15    &   0.21  &     0.21  \\
 NGC100      &      11.2  & 1.2  &   77.2  &   233  &    5.2  &     -23.81    &     96    &     2.07    &   0.15  &     0.16  \\
 NGC5023   &    4.8     &  0.8  &   78.4  &   160  &   2.9  &     -23.38    &     52    &     2.25    &   0.27  &     0.27  \\
 UGC5750   &     56.0       &  5.6  &   80.0  &   1171  &    46  &     -25.27    &     1125    &     1.56    &   -0.26  &     -0.18  \\
 UGC3371   &   12.8     &  3.1  &   82.0  &   681  &   20  &     -24.69    &     494    &     1.78    &   -0.09  &     -0.02  \\
 NGC4395    &    3.5  & 2.3  &   82.3  &   509  &  13  &     -24.40    &     312    &     1.89    &   -0.00  &    0.05  \\
 UGC11557  &    23.8  & 3.1  &   83.7  &   710  &    20  &     -24.67    &     520    &     1.80    &   -0.08  &    -0.01  \\
 UGC1230   &       51.0        &4.5  &   90.0  &   1278  &   34  &     -24.99    &     1027    &     1.71    &   -0.15  &     -0.07  \\
 ESO206-G014  &  60.3     &  5.2  & 91.3  &   1531  &    42  &     -25.12    &     1338    &     1.67    &   -0.19  &     -0.10  \\
 NGC2552   &      10.1    &  1.6  &   92.0  &   475  &    7.8  &     -23.97    &     213    &     2.09    &   0.14 &     0.18  \\
 UGC4278   &      10.5        &   2.3  &   92.6  &   691  &     13  &     -24.32    &     386    &     1.96    &   0.04  &     0.10  \\
 UGC634      &   35.0   & 3.1  &   95.1  &   984  &    20  &     -24.59    &     662    &     1.88    &   -0.03  &     0.05  \\
 ESO488-G049  &    23.0    & 4.4  &   95.3  &   1410  &   33  &     -24.92    &     1159    &     1.76    &   -0.13  &     -0.03  \\
 UGC5005     &     52.0    & 4.4  &   95.5  &   1406  &    33  &     -24.92    &     1153    &     1.77    &   -0.13  &     -0.03  \\
 UGC3137     &    18.4      & 2.0  &   97.7  &   669  &    11  &     -24.14    &     350    &     2.06    &   0.10  &     0.17  \\
 F574-1          &   96.0    & 4.5  &   99.0  &   1546  &    34  &     -24.91    &     1306    &     1.79    &   -0.12  &     -0.01  \\
 F568-3          &    77.0      & 4.0  &   100.5  &   1416  &    29  &     -24.78    &     1130    &     1.84    &   -0.08  &     0.02  \\
 ESO322-G019  &   45.2  & 2.5  &   100.7  &   878  &     14  &     -24.32    &     528    &     2.01    &   0.05  &     0.14  \\
 F563V2         &    61.0    & 2.1  &   101.3  &   755  &    11  &     -24.15    &     412    &     2.07    &   0.10  &     0.18  \\
 NGC 247       &     2.5  &   2.9  &   106.6  &   1156  &     18  &     -24.42    &     784    &     2.00    &   0.02  &     0.13  \\
 ESO444-G021  &    60.7  & 6.4  &   107.4  &   2603  &    56  &     -25.17    &     2760    &     1.75    &   -0.19  &     -0.05  \\
 F579V1         &     85.0   & 5.1  &   111.5  &   2223  &    40  &     -24.92    &     2134    &     1.85    &   -0.12  &     0.03  \\
 F568V1         &        80.0  & 3.2  &   115.8  &   1505  &     21  &     -24.44    &     1119    &     2.04    &   0.02  &     0.16  \\
 ESO374-G003  &   43.2      & 4.2  &   118.3  &   2084  &    31  &     -24.70    &     1856    &     1.97    &   -0.05  &     0.11  \\
 F568-1            &     85.0    &  5.3  &   130.0  &   4218  &    43  &     -25.13    &     1354    &     1.67    &   -0.03  &     -0.10  \\
 UGC628          &    65.0       &  4.7  &   130.0  &   3740  &    36  &     -25.02    &     1132    &     1.71    &    0.00  &     -0.07  \\
 UGC11616       &    72.8  & 4.9  &   133.2  &   4094  &    38  &     -25.04    &     1282    &     1.71    &   -0.00  &     -0.07  \\
 ESO186-G055  &   60.1      & 3.6  &   133.2  &   3041  &    25  &     -24.76    &     813    &     1.81    &   0.08  &     0.00  \\
 ESO323-G042  &    59.7     & 4.4  &   138.7  &   4020  &    33  &     -24.91    &     1221    &     1.78    &   0.04  &     -0.02  \\
 PGC37759       &     193.2       & 6.8  &   139.4  &   6195  &    60  &     -25.30    &     2318    &     1.65    &   -0.08  &     -0.12  \\
 ESO234-G013  &      60.9      & 3.7  &   139.4  &   3425  &    26  &     -24.74    &     949    &     1.84    &   0.08  &     0.02  \\
 F571-8              &   48.0          & 5.2  &   139.5  &   4765  &    42  &     -25.05    &     1577    &     1.73    &   -0.00  &     -0.05  \\
 F730V1             &    144.0     & 5.8  &   141.6  &   5523  &   49  &     -25.15    &     1953    &     1.71    &   -0.03  &     -0.07  \\
 UGC11648        &  46.7     & 3.8  &   142.2  &   3620  &    27  &     -24.74    &     1022    &     1.85    &   0.09  &     0.03  \\
 ESO215-G039  &    61.3  &  4.2  &   142.9  &   4037  &    31  &     -24.83    &     1208    &     1.82    &   0.06  &     0.01  \\
 ESO509-G091  &   72.8  & 3.7  &   146.8  &   3735  &     25  &     -24.68    &     1050    &     1.89    &   0.11  &     0.06  \\
 \hline  
\end{tabular}
\caption{Individual properties of LSBs. Columns: (1) galaxy name;  (2) distance; (3) disc scale length;  (4) optical velocity; (5) disc mass; 
(6) core radius; (7) central DM  density; (8) virial mass; (9) central surface density; (10) compactness of stellar mass distribution; (11) compactness of the DM mass distribution.}
\label{LSB_Tab4}
\end{table*}
\begin{table*}
\begin{tabular}{p{2.0cm}p{1.1cm}p{1.1cm}p{1.1cm}p{1.1cm}p{0.9cm}p{1.2cm}p{1.1cm}p{1.1cm}p{1.1cm}p{1.1cm}}
\hline
Name        &  $D$  &     $ R_d  $      &     $V_{opt}$        &         $M_d $              &    $ R_c $      &    $Log \, \rho _0$     &    $M_{vir}$          &       $Log \Sigma _0$           &       $Log C_* $        &       $Log C_{DM}$   \\
     \;            &   Mpc     &    kpc           &         km/s              &  $10^7 \,M_{\odot}$   &        kpc        &    $g/ cm^3$        &    $10^9 \,M_{\odot}$    &      $M_{\odot} / pc^2$        &                 \;                 &               \;                 \\
  (1)           &        (2)             &            (3)             &          (4)                     &         (5)          &           (6)                 &            (7)               &                  (8)                      &             (9)                 &              (10)           &   (11)   \\ 
\hline
ESO444-G047  &     62.4 & 2.7  &   148.4  &   2809  &   16  &     -24.38    &     662    &     2.01    &   0.19  &     0.13  \\
 UGC11454        &     92.1  & 4.5  &   150.3  &   4787  &    34  &     -24.85    &     1525    &     1.85    &   0.06  &     0.03  \\
 UGC5999          &    45.0     &  4.4  &   153.0  &   4851  &    33  &     -24.82    &     1540    &     1.87    &   0.07  &     0.04  \\
 UGC11819        &      59.2   & 5.3  &   154.6  &   6578  &    43  &     -25.10    &     1490    &     1.70    &   0.04  &     -0.08  \\
 ESO382-G006  &      65.4  &  2.3  &   160.0  &   3097  &   13  &     -24.29    &     449    &     2.01    &   0.27  &     0.13  \\
 ESO323-G073  &    69.6   &  2.1  &   165.3  &   2923  &    11  &     -24.14    &     398    &     2.08    &   0.32  &     0.18  \\
 NGC3347B       &    46.2   & 8.1  &   167.0  &   11760  &    78  &     -25.43    &     3369    &     1.63    &   -0.05  &     -0.14  \\
 ESO268-G044  &     49.9   & 1.9  &   175.6  &   3057  &   10  &     -24.01    &     406    &     2.16    &   0.36  &     0.23  \\
 ESO534-G020  &      226.7   & 16.7  &   216.6  &   40638  &    218  &     -25.86    &     17351    &     1.64    &   -0.17  &     -0.18  \\
 NGC7589         &      115.0 &  12.6  &   224.0  &   32831  &    146  &     -25.58    &     13657    &     1.75    &   -0.08  &     -0.07  \\
 UGC11748       &   73.1   & 3.1  &   240.7  &   9418  &    20  &     -24.22    &     1911    &     2.26    &   0.32  &     0.31  \\
 UGC2936         &     43.6   & 8.4  &   255.0  &   28363  &    82  &     -25.09    &     10784    &     1.99    &   0.07  &     0.12  \\
 F568-6             &     201.0     & 18.3  &   297.0  &   83839  &    249  &     -25.67    &     49173    &     1.89    &   -0.10  &     0.01  \\ 
\hline  
\end{tabular}
\caption{It continues from Tab. \ref{LSB_Tab4}.}
\label{LSB_Tab5}
\end{table*}

\section{Errors and scatter in the scaling relations}
\label{Error_Scaling_Relations}
\noindent   
In Tab. \ref{Error_Scaling_Relations_Tab}, the errors on the best fitting parameters of the scaling relations evaluated in this work  are shown.
The standard scatter $\sigma$ of individual galaxies data of the various scaling relations is also reported. In the 2D scaling relations, it is evaluated according to:
$\sigma = \sqrt{\sum_{i=1}^N   (y_i - f(x_i))^2 / N} $, where $N=72$, $y_i $ and $x_i $ are the logarithmic data in the $y$ and $x$ axes, respectively, and $f$ is the considered scaling function (a line).   
In the 3D scaling relations, the standard scatter  is evaluated according to:
$\sigma = \sqrt{\sum_{i=1}^N   (z_i - {\tilde{f}}(x_i,y_i))^2 / N} $, where $z_i$, $y_i $, $x_i $ are the logarithmic data in the $z$, $y$ and $x$ axes, respectively, and ${\tilde{f}}$ is the considered scaling function (a plane).   
 \begin{table}
\begin{tabular}{p{4.5cm}p{0.6cm}p{0.6cm}p{0.6cm}p{0.6cm}}
\hline
   Fitted relation       &   $ \Delta \, a $  &    $  \Delta  \, b$     &    $  \Delta  \, c$      &    $\sigma$    \\
   (1)                &                 (2)       &         (3)                  &                (4)        &     (5)     \\        
\hline       
Eq.\ref{Denormalisation_2}: $Log \,R_c (Log \, R_d) $     &      0.15            &          0.26               &        -      &      -   \\
Eq.\ref{MD_Vopt_fit}: $Log \,M_d (Log \,V_{opt}) $      &      0.25              &        0.12               &      -                  &     0.24    \\
Eq.\ref{rho0_Rc_fit}: $Log \, \rho_0 (Log \, R_c) $   &       0.07              &        0.05             &         -           &      0.21      \\
Eq.\ref{Rd_Vopt_fit}: $Log \,R_d (Log \,V_{opt}) $      &      0.25              &        0.13               &     -                   &     0.24    \\
Eq.\ref{Rd_Vopt_fit}: $Log \,R_c (Log \,V_{opt}) $      &     0.36             &        0.18               &        -                &     0.34    \\
Eq.\ref{Rd_Vopt_fit}: $Log \,\rho_0 (Log \,V_{opt}) $      &     0.56             &        0.28               &     -                   &     0.54   \\
Eq.\ref{Stellar_Compactness_1}: $Log \,R_d (Log \,M_d) $      &      0.23              &       0.02               &         -               &     0.16    \\
Eq.\ref{RD_Mvir_Cstar}: $Log \,M_d (Log \, V_{opt}, Log \,C_*) $     &      0.06               &          0.03               &          0.04       &    0.06    \\
Eq.\ref{RD_Mvir_Cstar}: $Log \,R_d (Log \,V_{opt}, Log \,C_*) $      &      0.02               &        0.01               &           0.02          &     0.02    \\
Eq.\ref{RD_Mvir_Cstar}: $Log \, R_c (Log \, V_{opt}, Log \,C_*) $   &       0.03              &         0.02              &           0.02            &     0.03     \\
Eq.\ref{RD_Mvir_Cstar}: $Log \, \rho_0 (Log \, V_{opt}, Log \,C_*) $      &        0.15              &       0.07                  &        0.10     &     0.13          \\
Eq.\ref{DM_Compactness_1}: $Log \,R_c (Log \,M_{vir}) $      &      0.26              &       0.02               &         -               &     0.15    \\
Eq.\ref{DM_Compactness_2}: $Log \, C_* (Log \,C_{DM}) $      &     0.01            &       0.06               &         -               &     0.15    \\
\hline
\end{tabular}
\caption{Errors and scatters on the various scaling relations. Columns: (1) relation; (2)-(3)-(4) error bars on the {\it first, second} and {\it third} (when present) fitting parameters; (5) standard scatter of the 72 individual galaxies data from the scaling relations.}
\label{Error_Scaling_Relations_Tab}
\end{table}
%

%\vfill\eject
\section{LSB rotation curves with their URC}
\label{LSB rotation curves with their URC}
\begin{figure*}
\begin{center}
\includegraphics[width=1\textwidth,angle=0,clip=true]{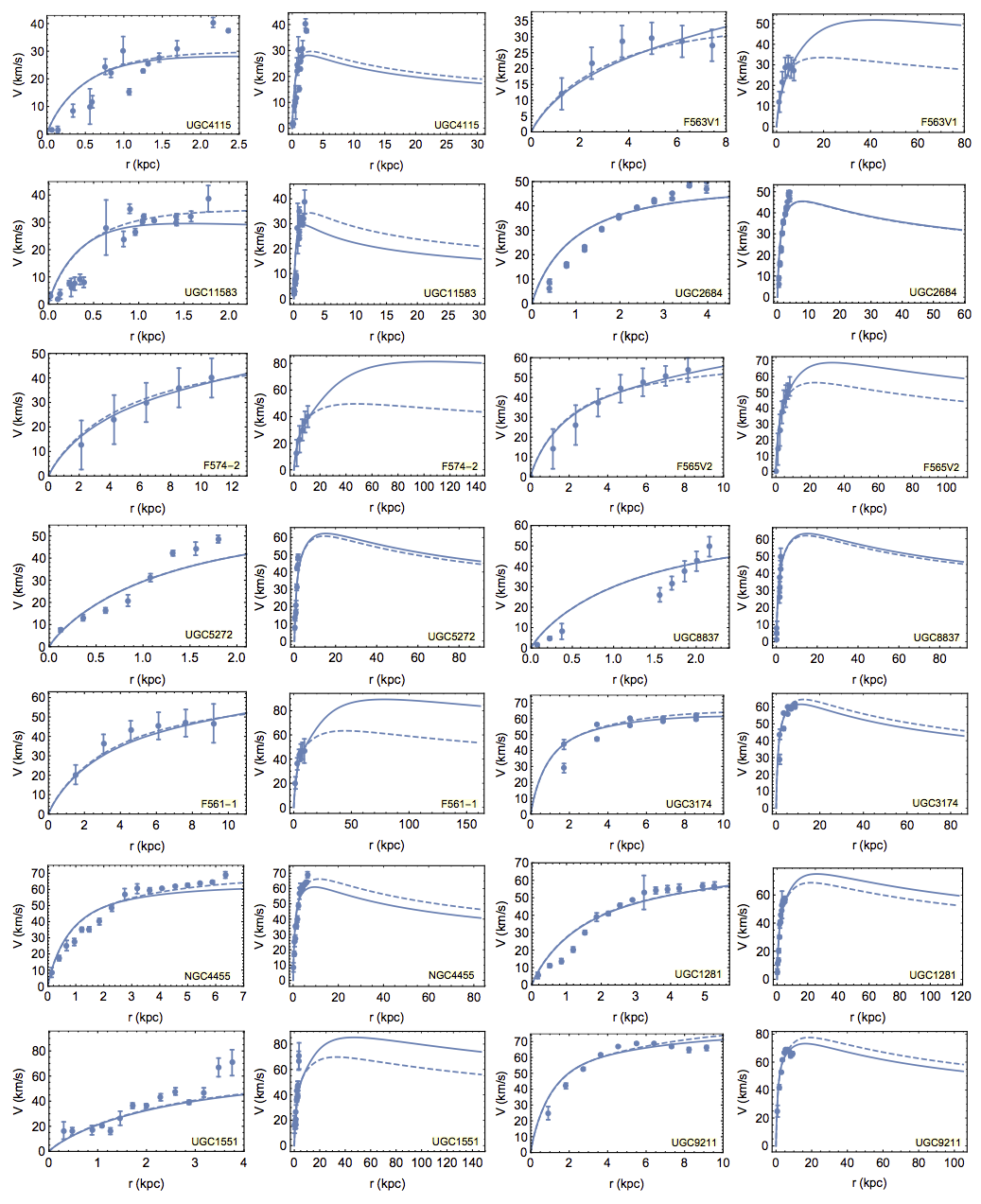}
\caption{LSBs rotation curves data with their URC given by Eq. \ref{Final_URC}-\ref{Final_URC_a}. The {\it solid} line is obtained for the $Log \, C_*$ values reported in Tab. \ref{LSB_Tab4}-\ref{LSB_Tab5} in Appendix \ref{Structural_properties} and is compared with the {\it dashed} line obtained for $Log \, C_*=0$. For each galaxy, we 
show the URC fit up to the farthest measurements ({\it left}) and up to the virial radius ({\it right}).}
\label{Single_URC}
\end{center}
\end{figure*}

\begin{figure*}
\begin{center}
\includegraphics[width=1.0\textwidth,angle=0,clip=true]{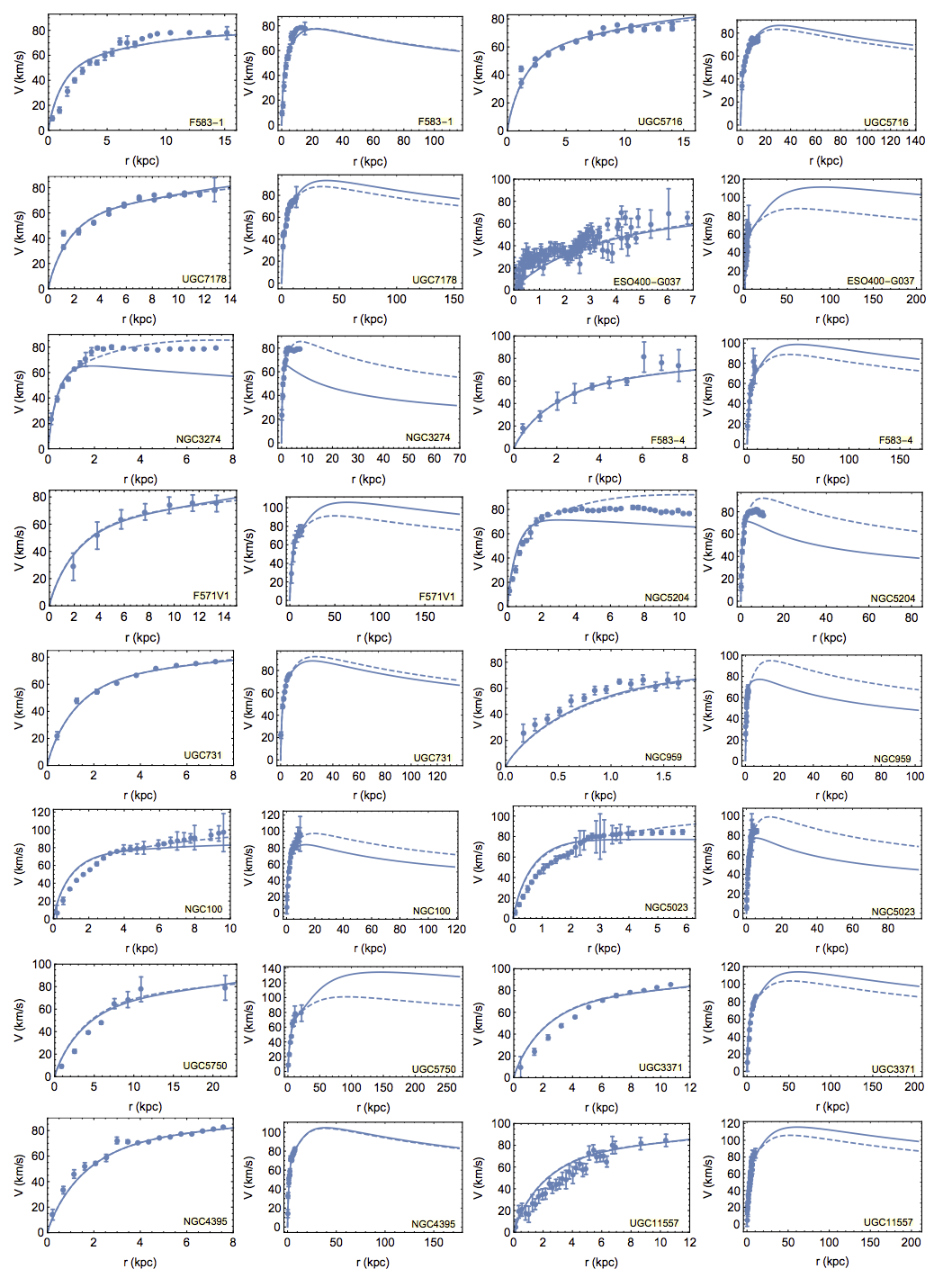}
\caption{It continues from Tab. \ref{Single_URC}.}
\label{Single_URC_1}
\end{center}
\end{figure*}

\begin{figure*}
\begin{center}
\includegraphics[width=1\textwidth,angle=0,clip=true]{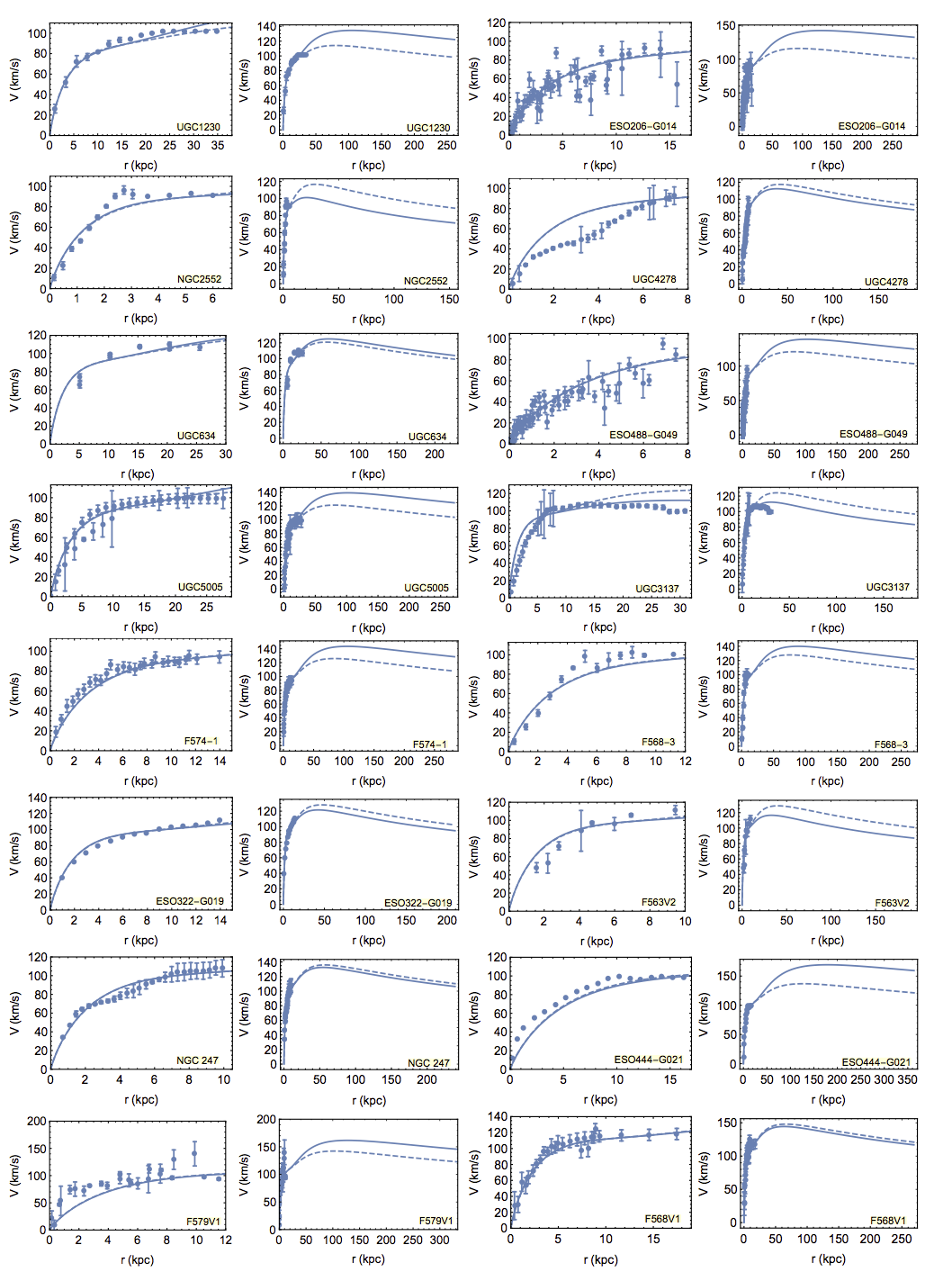}
\caption{It continues from Tab. \ref{Single_URC_1}.}
\label{Single_URC_2}
\end{center}
\end{figure*}

\begin{figure*}
\begin{center}
\includegraphics[width=1\textwidth,angle=0,clip=true]{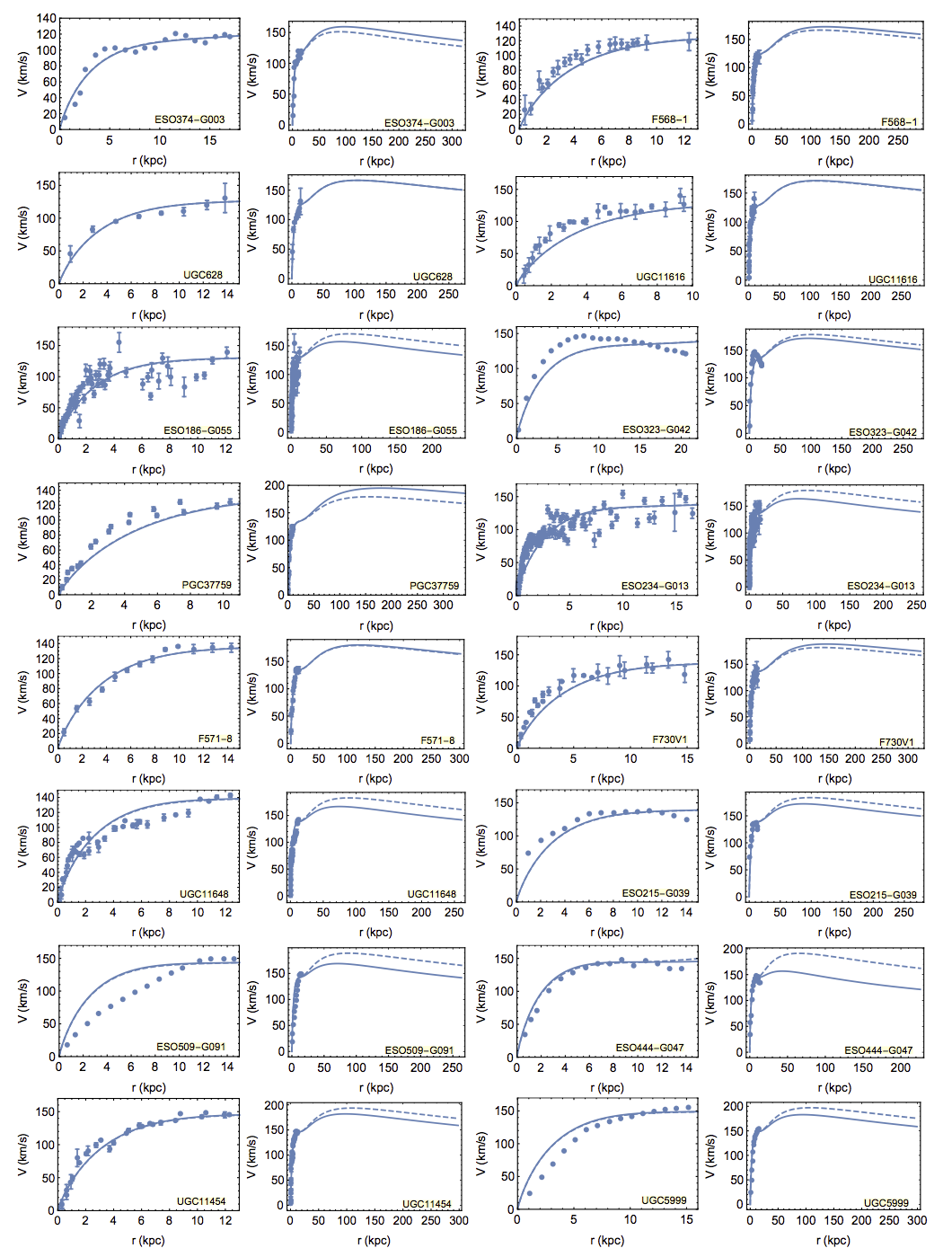}
\caption{It continues from Tab. \ref{Single_URC_2}.}
\label{Single_URC_3}
\end{center}
\end{figure*}

\begin{figure*}
\begin{center}
\includegraphics[width=1\textwidth,angle=0,clip=true]{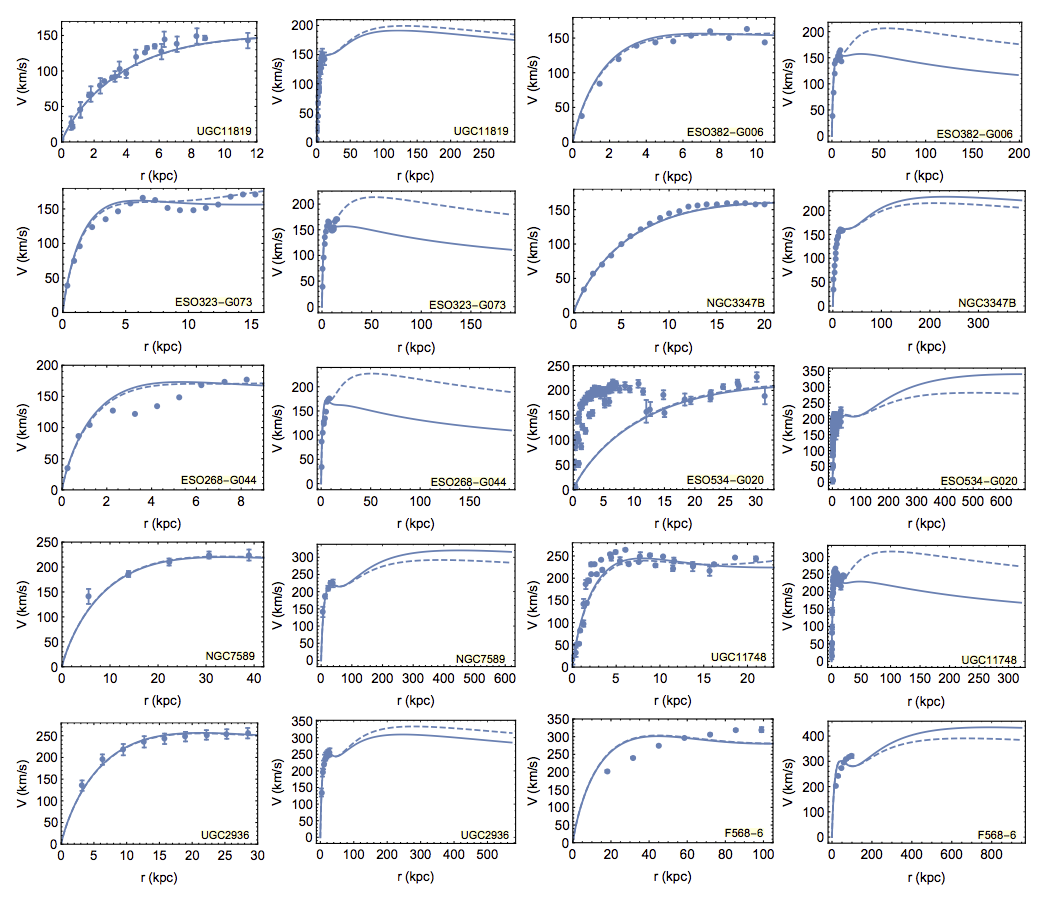}
\caption{It continues from Tab. \ref{Single_URC_3}.}
\label{Single_URC_4}
\end{center}
\end{figure*}

\noindent
We show in Fig. \ref{Single_URC}-\ref{Single_URC_1}-\ref{Single_URC_2}-\ref{Single_URC_3}-\ref{Single_URC_4} the LSBs  rotation curves data together with their URC, taking into account Eq. \ref{Final_URC}-\ref{Final_URC_a} and the values of 
$R_{opt} \equiv 3.2 \,R_d$, $V_{opt}$ and $C_*$ reported in Tab. 
\ref{LSB_Tab4}-\ref{LSB_Tab5} in Appendix \ref{Structural_properties}. We also show the URC for the case $Log \, C_* = 0$ in Fig. \ref{Single_URC}-\ref{Single_URC_1}-\ref{Single_URC_2}-\ref{Single_URC_3}-\ref{Single_URC_4}. In comparing the URC model with the 72 individual rotation curves, in 21 of them we have assumed a random systematic error running from $\simeq 3\%$ to $\simeq 16\%$ in their amplitudes (velocity measurements). In Tab. \ref{LSB_Modified_1}, the changes applied are specified. Removing such systematics improves fits which were already successful.    
Let us stress that the URC can  help determining  how well an individual RC correctly reflects the mass distribution of the galaxy.  

\begin{table}
\begin{tabular}{p{2.0cm}p{1.0cm}}
\hline
Name         &     $ f \% $       \\
   (1)           &        (2)               \\ 
\hline
UGC2684  &   + 10.9   \\
F565V2      &   + 8.8   \\
F561-1        &  -  7.9    \\
UGC3174    &  - 9.7    \\
UGC1551     & - 14.3  \\
UGC9211     &  + 4.8    \\
F583-1          &  - 8.1    \\
ESO400-G037  & - 7.1    \\
NGC959        &  - 15.9   \\
F574-1           & - 8.1     \\
ESO444-G021  & - 9.3   \\
F579V1       &   - 16.1   \\
ESO374-G003  & - 5.9   \\
F568-1     &  - 9.2    \\
UGC11616    &  - 7.5  \\
PGC37759    & - 10.8   \\
F730V1     &  - 9.2    \\
ESO215-G039   &    - 10.5   \\
UGC11454      &  -  3.3     \\
NGC3347B   &  - 6.0  \\
ESO268-G044 &  - 5.7   \\
\hline  
\end{tabular}
\caption{List of galaxies in which we have left the amplitude of the RC to freely vary by $f \%$. Columns: (1) galaxy name; (2) correction to the velocity values of the RCs data, expressed in $f \%$.}
\label{LSB_Modified_1}
\end{table}

\end{appendix}

%\bibliography{Bibliografia}

\bsp	% typesetting comment
\label{lastpage}
\end{document}